\documentclass[final,3p,times]{elsarticle}

\graphicspath{{figs/}}
\usepackage{epstopdf}
\usepackage{url,hyperref,microtype,subcaption}
\usepackage{nccmath}
\usepackage{amsfonts,latexsym,eucal,amsmath,amsthm,amssymb,bbm} 
\usepackage{graphicx}

\usepackage{xcolor}
\usepackage{amsbsy}
\usepackage{bm}       
\usepackage{dsfont}
\usepackage[final]{changes}
\usepackage{indentfirst}
\usepackage{multirow}
\usepackage{pdfpages}
\usepackage{txfonts}
\usepackage{url}

\usepackage{dblfloatfix}
\usepackage{nicefrac}       
\usepackage{xfrac}
\usepackage{xr}
\usepackage{cleveref}

\def\ve{\varepsilon}

\journal{Immunological Methods}

\begin{document}

\begin{frontmatter}



\title{CEI: A Clonal Expansion Identifier for T-cell receptor clones
  following SARS-CoV-2 vaccination}

\author[a1]{Yunbei Pan}
\author[a2]{Christian Hofmann}
\author[a3]{Barbara Banbury}
\author[a3]{Harsh Patel}
\author[a3]{Stephanie A. Bien}
\author[a1,a4]{Tom Chou}
\author[a2]{Otto O. Yang}
\affiliation[a1]{organization={Department of Computational Medicine, UCLA},
            addressline={621 Charles E. Young Drive S.}, 
            city={Los Angeles},
            postcode={90095-1766}, 
            state={CA},
            country={USA}}

\affiliation[a2]{organization={Depts. of Medicine and Microbiology, Immunology,
  and Molecular Genetics, UCLA},
            addressline={615 Charles E. Young Drive S., East},
            city={Los Angeles},
            postcode={90095}, 
            state={CA},
            country={USA}}

\affiliation[a3]{organization={Adaptive Biotechnologies},
  addressline={1165 Eastlake Ave E},
              city={Seattle},
            postcode={98109}, 
            state={WA},
            country={USA}}

\affiliation[a4]{organization={Department of Mathematics, UCLA},
            addressline={520 Portola Plaza, MS 6363},
              city={Los Angeles},
            postcode={90095-1555}, 
            state={CA},
            country={USA}}







\begin{abstract}
Each T cell typically carries a specific T-cell receptor (TCR) that
determines its specificity against an epitope presented by the HLA
complex on a target cell. Antigenic challenge triggers the expansion
of reactive cells within a diverse pool of T cells with randomly
generated receptors, a process that results in epitope-driven shifts
of TCR frequencies over time. Here, we analyze the effects of
SARS-CoV-2 vaccination on the TCR populations in peripheral blood
drawn from seven COVID-naive individuals, before vaccines were widely
available.  To identify SARS-CoV-2 vaccine-associated TCR sequences
among the $\sim 10^{5}-10^{6}$ TCR sequences sampled before and after
vaccination, we develop statistical criteria to detect significant
increases in abundance of positive TCR clones. Application of our
statistical methods shows a robust identification of TCR sequences
that respond to SARS-CoV-2 vaccination \textit{in vivo}, illustrating
the feasibility of quantifying the clone-specific dynamics of T-cell
abundance changes following immunological perturbations.
\end{abstract}


%

\end{frontmatter}

\section*{Introduction}

mRNA vaccines that deliver SARS-CoV-2 spike protein (S protein) to the
blood have been highly effective in reducing COVID-19 morbidity and
mortality
\citep{Baden2021mrna,garcia2021multiple,goel2021mrna,ibarrondo2021primary,link2023estimation,polack2020safety,zhang2022humoral}. Vaccination
promotes HLA-restricted T-cell responses against the S protein that
play a critical role in this protective effect. These responses are
mediated by TCRs recognizing epitopes from the S protein. TCRs are
heterodimers
\citep{augusto2023common,buckley2022hla,sette2023t,taus2023persistent,taus2022dominant}
consisting of an $\alpha$-chain (TCRA) and a $\beta$-chain (TCRB)
\citep{murphy2016janeway}. In the generation of T cells, somatic
recombination of the different domains [the variable (V) and joining
  (J) domains in TRA and TRB, and the diversity (D) segment in TRB] in
the associated TCR sequence occurs. Along with nucleotide insertions
and deletions, a complementarity-determining region 3 (CDR3) is
generated with up to $10^{15}$ possible combinations of TCR sequence
\citep{murphy2016janeway,nikolich2004many,sewell2012must,arstila1999direct}.

Of this possible diversity, it has been estimated that $\sim 10^{8}$
unique TCR sequences are circulating in the peripheral blood of a
person
\citep{lythe2016many,mora2019many,qi2014diversity,warren2011exhaustive,
  jenkins2009composition}.  CTLs (cytotoxic T lymphocytes) are
CD8$^{+}$ T lymphocytes that eliminate target cells through cytolysis
and apoptosis. A second group of CD4$^+$ lymphocytes includes key
``helper T cells'' that direct immune responses by supporting CTLs and
contributing to antigen-specific antibody production.

These T-cell responses have been shown to play a key role in
mitigating the pathogenicity of SARS-CoV-2
infection~\citep{murphy2016janeway,aoki2024cd8+}, and an early T-cell
response has been associated with a mild outcome of COVID-19
\citep{mallajosyula2021cd8+,swadling2022pre}. In contrast to rapidly
waning antibody responses \citep{ibarrondo2021primary,
  ibarrondo2020rapid}, SARS-CoV-2-specific T cells persist over longer
periods and provide an immediate immune response against COVID-19
re-infection \citep{rodda2021functional,wheatley2020evolution}.
Furthermore, the evolution of SARS-CoV-2 into variants has been marked
predominantly by mutations in and around the S protein receptor
binding domain (RBD), leading to escape from neutralizing antibodies
(which generally target the RBD), while T cells target epitopes
distributed throughout the S protein and thus have a broader
recognition of variants.
%
%
The large size and abundance of the S protein make it a good target
for the adaptive immune response \citep{grifoni2020targets}. To better
understand the T-cell response against COVID-19, large libraries of
TCR sequences have been analyzed using various bioengineering methods
\citep{jokinen2023tcrconv,ferretti2020unbiased,grifoni2021sars}.

Quantifiable detection of vaccine-induced peptide-specific expansion
of TCR clones is required in many applications. However, challenges
include analyzing $\sim 10^6$ TCR clones without replicates and
defining abundance changes among clones with highly dispersed
abundances.  In addition, different subjects may experience very
different immune response intensities. These factors complicate the
accurate assessment of T-cell clone responses during the
post-vaccination visits and the capture of significant changes in the
abundances of individual clones.

Several advanced Bayesian methods for detecting clonal expansions,
such as NoisET \cite{koraichi2022noiset} and edgeR
\citep{chen2018differential}, which are based on Bayesian models, and
have been recently developed. The incomplete overlap in identified
clones using these methods and the absence of a ground truth for
evaluating their reliability motivates us to develop a new method
grounded in fundamental statistical principles for a better
comparative analysis.

In this study, we develop CEI, a Python package designed for the
rapid, scalable detection of clonal expansion from single samples
from repertoires pre- and post-vaccination (or in general, any
perturbation). Applied to TCRB repertoires from blood drawn before and
after mRNA/adenoviral-vectored vaccination, CEI identifies
vaccine-associated TCRB sequences using a two-proportion test and
analyses of differences and log-fold change of abundances.  While
developed for TCRB data in this work, the framework is applicable to
other replicate-limited count comparisons (e.g., TCR/BCR repertoires
or CRISPR screens) where feature-level proportions are contrasted
across two distinct conditions.

\section*{Materials and Methods}

\subsection*{Data preprocessing}

Relevant to this study (comparing pre- and post-vaccination TCR
abundances in COVID-naive subjects), samples were collected from seven
individuals (labeled CLE0083, CLE0099, CLE0100, CLE0101, CLE0108,
CLE0117, and CLE0136) between April 2020 and May
2021. Post-vaccination samples were taken from subjects in
January-September 2021, very early in the pandemic before vaccines
were widely available. These subjects received their first
vaccinations between 12/18/2020 and 1/8/2021, and their second
vaccinations between 01/06/2021 and 01/30/2021. Subject CLE0099
received the AstraZeneca chimpanzee adenovirus vectored vaccine while
all others received the BNT162b2 mRNA vaccine.  Additionally, all
subjects had no prior history of COVID-19; baseline spike RBD
antibodies were also confirmed negative by ELISA around the time of
the first vaccination. This dataset represents a rare and
scientifically valuable window into the earliest immune responses to
SARS-CoV-2 vaccination, captured during a critical period at the onset
of the global vaccine rollout.  This temporal isolation from natural
infection ensures that the observed TCR dynamics are most likely to be
attributable to vaccination.

Overall, samples were taken at three time points: baseline (0-3 days
before the first vaccination, labeled B), 14 to 20 days after the
first vaccination but before (labeled P1), and 11 to 21 days after the
second vaccination (labeled P2). Further details are provided in Table
~\ref{tab:individuals}. The TCRs of both CD4$^+$ and CD8$^+$ T cells
in these samples were sequenced by Adaptive Biotechnologies.

\begin{table*}[htp]
\centering
\renewcommand*{\arraystretch}{1.3}
    \begin{tabular}{cccccc|ccc}
    \hline
    \multirow{2}{*}{Subjects} & \multicolumn{5}{|c|}{Demographics} & \multicolumn{3}{c}{Sample times versus vaccine doses (days)}  \\ \cline{2-9} 
    \: & \multicolumn{1}{|c|}{Age} & \multicolumn{1}{c|}{Sex} & \multicolumn{1}{c|}{Race} & \multicolumn{1}{c|}{Hisp} & \multicolumn{1}{c|}{Vaccine} & \multicolumn{1}{c|}{B to Vacc $\#$1} & \multicolumn{1}{c|}{P1 to Vacc $\#$1}  & \multicolumn{1}{c}{P2 to Vacc $\#$2} 
    \\ \hline
    CLE0083 & \multicolumn{1}{|c|}{49} & \multicolumn{1}{c|}{M} & \multicolumn{1}{c|}{A} & \multicolumn{1}{c|}{N} & \multicolumn{1}{c|}{BNT162b2} & \multicolumn{1}{c|}{-3} & \multicolumn{1}{c|}{18} & \multicolumn{1}{c}{18}
    \\ \hline
    CLE0099 & \multicolumn{1}{|c|}{51} & \multicolumn{1}{c|}{M} & \multicolumn{1}{c|}{W} & \multicolumn{1}{c|}{N} & \multicolumn{1}{c|}{AZD1222} & \multicolumn{1}{c|}{0} & \multicolumn{1}{c|}{15} & \multicolumn{1}{c}{14}
    \\ \hline
    CLE0100 & \multicolumn{1}{|c|}{55} & \multicolumn{1}{c|}{M} & \multicolumn{1}{c|}{A} & \multicolumn{1}{c|}{N} & \multicolumn{1}{c|}{BNT162b2} & \multicolumn{1}{c|}{-2} & \multicolumn{1}{c|}{16} & \multicolumn{1}{c}{14}
    \\ \hline
    CLE0101 & \multicolumn{1}{|c|}{36} & \multicolumn{1}{c|}{F} & \multicolumn{1}{c|}{A} & \multicolumn{1}{c|}{N} & \multicolumn{1}{c|}{BNT162b2} & \multicolumn{1}{c|}{0} & \multicolumn{1}{c|}{14} & \multicolumn{1}{c}{13}
    \\ \hline
    CLE0108 & \multicolumn{1}{|c|}{50} & \multicolumn{1}{c|}{F} & \multicolumn{1}{c|}{Iranian} & \multicolumn{1}{c|}{N} & \multicolumn{1}{c|}{BNT162b2} & \multicolumn{1}{c|}{-1} & \multicolumn{1}{c|}{20} & \multicolumn{1}{c}{11}
    \\ \hline
    CLE0117 & \multicolumn{1}{|c|}{64} & \multicolumn{1}{c|}{M} & \multicolumn{1}{c|}{W} & \multicolumn{1}{c|}{N} & \multicolumn{1}{c|}{BNT162b2} & \multicolumn{1}{c|}{0} & \multicolumn{1}{c|}{14} & \multicolumn{1}{c}{15}
    \\ \hline
    CLE0136 & \multicolumn{1}{|c|}{42} & \multicolumn{1}{c|}{F} & \multicolumn{1}{c|}{W} & \multicolumn{1}{c|}{N} & \multicolumn{1}{c|}{BNT162b2} & \multicolumn{1}{c|}{0} & \multicolumn{1}{c|}{18} & \multicolumn{1}{c}{21}
    \\ \hline
    \multicolumn{6}{c|}{Mean (Range)} & \multicolumn{1}{c|}{-1 (-3 to 0)} & \multicolumn{1}{c|}{16 (14 to 20)} & \multicolumn{1}{c}{15 (11 to 21)}
    \\ \hline
\end{tabular}
\caption{Vaccination records for all the individuals. BNT162b2 is an
  mRNA vaccine, while AZD1222 is a chimpanzee adenovirus-vectored
  vaccine. ``Vacc $\#$1'' and ``Vacc $\#$2'' refer to the first and
  second doses of the vaccine, respectively. The timing of sample
  collection in relation to vaccination is denoted by sample times
  versus vaccine doses. For example, in ``B to Vacc $\#$1,'' -3
  represents a pre-vaccination baseline sample that was taken three
  days before the first ''P1'' vaccination.}
\label{tab:individuals}
\end{table*}
%



\subsection*{Post processing of immunosequencing data for
    differential abundance analysis}

Differential abundance analysis is a statistical framework used to
determine whether a particular receptor rearrangement is more abundant
in one sample than in another. It can be applied to nucleotide, amino
acid, or bioidentity (V gene/amino acid/J gene) sequences depending on
the desired resolution and biological context. While nucleotide-level
matching provides the highest resolution, amino acid-level matching
allows for grouping of clonotypes that may differ at the nucleotide
level but encode the same protein sequence. We employed the IMSEQ
\citep{Kuchenbecker2015IMSEQ} algorithm to collapse nucleotide
sequences for identical clonotypes for downstream differential
abundance analysis of bioidentities. Since the amino acid sequence
determines protein function, we established criteria to determine
whether two nucleotide sequences are identical:

\begin{itemize}
    \item If the V subtype, CDR3 amino acids, and J subtype are all
      identical, we consider these two sequences to be identical;

    \item If the two sequences differ only in their V subtype, and one
      or both are ambiguous, we treat them as the same if the longer
      sequence contains the shorter;

    \item If the two sequences differ by only one base (A, C, T, G) in
      the nucleotide sequence of the CDR3 region, we consider them
      identical.
\end{itemize}
For identical sequences, we sum their template counts for the
subsequent analysis of changes in abundance.

\subsection*{Quantification of clone abundance changes}

Due to the specificity of TCR binding to assigned antigen peptides, we
can statistically screen for sequences that expand after vaccination
by comparing with the baseline sample. Our analysis identifies TCR
sequences that are enriched in response to vaccine-associated antigens
(including but not limited to S protein) as ``positive'' and those
that do not target S proteins as ``negative.'' We assume that all T
cells and SARS-CoV-2 protein-expressing cells generated by mRNA
vaccination are well-mixed, and neither the exposure nor the expansion
capacity of reactive T cells is limiting. Although false-positive
sequences (\textit{i.e.}, non–vaccine-associated sequences that
expand) may occur, their number is relatively small compared to the
true-positive sequences. Likewise, the likelihood of false-negative
sequences (\textit{i.e.}, SARS-CoV-2–specific sequences not
responding) can be neglected.

Intuitively, we need to assess the magnitude of the difference in the
abundances of individual clones (here, a clone denotes the collection
of cells with the same V gene, CDR3 amino acid, and J gene identity),
measured in samples taken before (baseline) and after vaccination. In
addition, we can evaluate the intensity of the response at each
post-vaccination time point.  One approach is to treat the clone
abundances across clone identities as a distribution and then measure
the dissimilarity between the distributions at different visit times
within one individual. For instance, we can consider metrics/distances
such as KL divergence \citep{KL1951div}, Wasserstein distance
\citep{Kantorovich1960mathematical,Vaserstein1969markov}, cosine
similarity \citep{Singhal2001modern}, Bhattacharyya distance
\citep{Bhattacharyya1946measure}, among others. However, these
approaches face challenges in effectively addressing scenarios where
clones were not initially sampled at the baseline but have grown
significantly after vaccination. Moreover, because these metrics
typically quantify changes in the global shape of the distributions,
very high-abundance clones may dominate the metrics and thereby
obscure relatively small clones that have nonetheless undergone
substantial growth. It is also probable that extremely large clones
are not SARS-CoV-2 specific if they did not appreciably expand upon
vaccination.

Here, we will focus on the changes in the abundance of individual
clones between samples taken from two repertoires. We assign a fixed
index $1\leq i \leq I$ to each clone, where $I$ is the total number of
different clones observed in the two samples drawn from the same
subject before (B) and after (P) vaccination. Given a clone with index
$i$, we denote its true \textit{in vivo}, whole body sizes at time
$t\in \{B, P\}$ as integers $n_{i}^{(t)}$. We also define the
corresponding total number of T cells at these two time points as
$N^{(B)}$ and $N^{(P)}$, respectively.

Assuming $N^{(B)} \approx N^{(P)}$, consider a scenario where the
abundance of a sampled clone changes from $10,000$ to
$10,100$. Although the absolute difference of $100$ is appreciable, if
this difference arises in an already-large clone, the relative change
in abundance could be small and attributed to the sampling variability
or intrinsic fluctuations in a ``homeostatic'' large clone.  On the
other hand, a fold-change from $0$ to $5$ is considered large
(infinite in this case), but an absolute difference of $5$ might also
arise from noise, sampling or otherwise.  Therefore, we will use both
the absolute difference and the fold-change as quantities for the
identification of statistically significant changes in abundance.

We motivate and develop two clone-level quantities or ``indicators''
that can be used to effectively screen expanded clones and measure the
strength of response following a perturbation, vaccination in our
case.
\begin{enumerate}
    \item {\bf Difference in clone abundance} The normalized change in
      abundances from data is defined as
    \begin{equation}
        d_i \equiv  \hat{p}_i^{(P)} - \hat{p}_{i}^{(B)} 
    \end{equation}
where $\hat{p}_{i}^{(t)} = \frac{s_i^{(t)}}{S^{(t)}}$ represents the
abundance fraction of clone $i$ in the sample drawn at time $t\in\{B,
P\}$. \\

\item {\bf Regularized log-fold change} We define the log of a
  regularized abundance ratio by
    \begin{equation}
        r_i \equiv \log_{2}\frac{s_{i}^{(P)}+\ve}{S^{(P)}} - \log_{2}\frac{s_{i}^{(B)}+\ve}{S^{(B)}}
    \end{equation}
where $\ve$ is a Haldane-Anscombe correction to stabilize zero counts
(the default value $\ve=0.5$ is typically used). Taking the logarithm
defines $r_i$ over the entire real line ($-\infty < r_{i} < \infty$).
\end{enumerate}
%

\subsubsection*{Variance, Wald statistic, and $p$-values for $d_i$}

Under a simple random sampling assumption (\textit{i.e.,} each T cell
has the same probability of being sampled) with total sample size
$S^{(B)}$ and $S^{(P)}$, the abundances $\left\{s_{i}^{(t)}\right\}$
of sampled clones $i$ at time $t\in\{B, P\}$ follow a multinomial
distribution:
\begin{equation}
  \left(s_{1}^{(t)}, s_{2}^{(t)}, \ldots, s_{I}^{(t)} \right) \sim
  \text{Multinomial}\left(S^{(t)};\;p_{1}^{(t)}, p_{2}^{(t)}, \ldots,
  p_{I}^{(t)} \right), \qquad t\in \{B,P\}
  \label{MULTINOMIAL}
\end{equation}
where $p_{i}^{(t)} = \frac{n_{i}^{(t)}}{N^{(t)}}$ is the probability
of the clone $i$ being sampled at time $t$. Note that because very few
clones from the entire body are sampled, $S^{(t)}=\sum_{i=1}^{I}
s_{i}^{(t)} \ll N^{(t)}$ for $t\in \{B, P\}$ and $\sum_{i=1}^{I}
p_{i}^{(t)} \ll 1$.

Since only a small number of T cell clones respond to vaccination, we
consider no change in abundance, \(p_{i}^{(B)} = p_{i}^{(P)} = p_i\),
as the null hypothesis ($H_{0}$). We expect very few clones to violate
this hypothesis.

Since the joint statistical properties for two signals (difference and
log-fold change in abundance) under a true multinomial null are
analytically complex, we adopt a resampling-based approach to assess
clone abundance change significance. To reduce computational burden,
we replace the full multinomial in Eq.~\ref{MULTINOMIAL} with a
marginalized probability distribution, treating each clone
independently. Specifically, considering the probability of each clone
to be sampled is small and library sizes are extremely large, we use a
Poisson model:
\begin{equation}
  s_{i}^{(t)} \sim \text{Poisson}\left(\lambda_i^{(t)}\right), \qquad
  \text{where } \;\lambda_{i}^{(t)} = S^{(t)} p_i \;\text{ and }\;
  t\in\{B, P\}.
\end{equation}
%
Since the true $p_i$ is unknown, we approximate it using the
pooled proportion estimator:
\begin{equation}
    \hat{p}_i = \frac{s_{i}^{(B)} + s_{i}^{(P)}}{S^{(B)}+S^{(P)}}
\end{equation}
The variance of the difference $d_i$ for clone $i$ is thus calculated by
\begin{equation}
  \mathrm{Var}(d_i) = \mathrm{Var}\left( \frac{s_i^{(P)}}{S^{(P)}} \right) + \mathrm{Var}\left(\frac{s_i^{(B)}}{S^{(B)}}\right) =  \frac{\hat{p}_{i}}{S^{(P)}} + \frac{\hat{p}_{i}}{S^{(B)}}
\end{equation}

To better balance the effects of the difference and log-fold change,
we need to nondimensionalize the difference. Consider taking a sample
from each clone as a single experiment. Instead of focusing on global
$Z$ scores, we should utilize Wald statistics to find asymptotic $Z$
scores under $H_0$ for each clone separately. Therefore, for clone
$i$, we have
\begin{equation}
  Z_{d_i} = \frac{d_i}{\sqrt{\hat{p}_{i}\left(\frac{1}{S^{(P)}}
      + \frac{1}{S^{(B)}}\right)}}
\end{equation}
The one-sided p-value $p_{d_i}=1-\Phi(Z_{d_i})$ for the clone $i$ is
used to detect expansion. Here $\Phi(x) =
\mfrac{1}{2}\big[1+\mathrm{erf}\big(x/\sqrt{2}\big)\big]$ is the
cumulative density function of the standard normal distribution.

\subsubsection*{Variance, Wald statistic, and $p$-values for
 log-fold change $r_i$}

Considering the empirical abundance fraction $s_{i}^{(t)}/S^{(P)}$ of clone $i$ in sample $t \in \{B, P\}$. As
assumed previously, the count of each clone follows a Poisson
distribution. Given the definition of the log-fold change
$r_i=\log_{2} {\tilde p}_{i}^{(P)} - \log_2 {\tilde p}_{i}^{(B)}$, the
null hypothesis for this case is supposed to satisfy $H_0: r_i = 0$
(no change upon vaccination). By assuming the independence between
visits, the multivariate delta method yields the following variance:
\begin{equation}
    \begin{aligned}
      \operatorname{Var}(r_i) \approx
      \frac{1}{(\ln 2)^2}\left(\frac{1}{{\tilde p}_{i}^{(P)} S^{(P)}}
      + \frac{1}{{\tilde p}_{i}^{(B)} S^{(B)}}\right)
    \end{aligned}
\end{equation}
where ${\tilde p}_{i}^{(t)} = \frac{s_{i}^{(t)}+\ve}{S^{(t)}}$ ($t \in
\{B, P\}$) are empirical proportions smoothed by a Haldane-Anscombe
correction $\ve$ (default $\ve=0.5$).

Applying the Wald test gives us the asymptotic $Z$ scores:
\begin{equation}
  Z_{r_i} = \frac{r_i}{\sqrt{{\rm Var}(r_i)}} =
  \frac{\ln2 \left(\log_{2} {\tilde p}_{i}^{(P)} - \log_2 {\tilde p}_{i}^{(B)}\right)}{\sqrt{\frac{1}{{\tilde p}_{i}^{(P)} S^{(P)}}
      + \frac{1}{{\tilde p}_{i}^{(B)} S^{(B)}}}}
\end{equation}
which is approximately normally distributed when $H_0$ holds. In this
case, the one-sided p-value is found from $p_{r_i} = 1 -
\Phi(Z_{r_i})$.

\subsubsection*{CEI-Joint method}

Since both differences in abundance and log-fold changes between two
samples are considered important in identifying amplified clones, it
is natural to consider their associated $Z$ scores as components of a
two-dimensional vector $Z_i = (Z_{d_i}, Z_{r_i})^{\top}$.

To include standardized statistics for both differences and log-fold
changes into a single test statistic, we can consider a score based on
a modified Mahalanobis distance for each clone $i$,
\begin{equation}
    D_i^2 \equiv (Z_i - m)^{\top} \; \Sigma^{-1} \;(Z_i-m)
\end{equation}
where $\Sigma$ denotes a covariance matrix
\begin{equation}
    \Sigma = {\mathbb{E}}\left[(Z_{d_i} - m_{d})(Z_{r_i}-m_{r}) \right]
\end{equation}
Here, we use the trimmed median vector $m$ after removing the largest
and smallest $10\%$ of clones by $\|Z_{i}\|_{\infty}$. Vaccine
stimulation leads to a very small number of clones experiencing
significant expansion, induces modest contraction in several others,
but leaves the majority of clones statistically
unchanged. Consequently, the median serves as a more stable
representation of the ``no-change'' baseline than the mean.

We control false positives by testing our CEI-Joint method with a
multinomial permutation under the no-change null hypothesis.  By
applying the pooled clone frequencies $\hat{p}_i$, we redraw entire
libraries as $s_{i,r}^{(B)}\sim \mathrm{Multinomial}(S^{(B)},\hat
p_{i})$ and $s_{i,r}^{(P)}\sim \mathrm{Multinomial}(S^{(P)},\hat p)$
in each permutation $r = 1, \cdots, R$. For each permutation, we
recompute the same per-clone statistics (\textit{i.e.}, $Z_{d_i}$,
$Z_{r_i}$) and the 2D Mahalanobis distance. We then pool all permuted
distances across clones to obtain the null distribution. The empirical
$p$-value for clone $i$ is calculated as follows:
\begin{equation}
  \hat p_i=\frac{1+\#\{D^2_{\text{null}}\ge D^2_{i}\}}{1+IR}
  \qquad\qquad\text{ (with $I$ clones, $R$ permutations)}
\end{equation}
This formula avoids zero $p$-values and accounts for dependence
between the two Z scores. Finally, we apply the Benjamini-Hochberg
(BH) procedure to control the false discovery rate (FDR)
\cite{dewitt,stats}. By setting an FDR threshold, denoted as $\alpha$
(\textit{e.g.}, $\alpha = 10\%$). Given a total of $I$ clones and
their corresponding $p$-values, denoted as $p_{1}, p_{2}, \ldots,
p_{I}$, we sort these $p$-values in ascending order to obtain
$p_{r_1}, p_{r_2}, \ldots, p_{r_I}$. Next, we identify the largest
index $i$ such that $p_{r_i} \leq (i/I) \alpha$. The clones ranked
from $1$ to $i$ are referred to as ``expanded.'' This filtering step
preferentially removes clones whose apparent changes are consistent
with random sampling noise, given the fixed library sizes.


%
%

\subsubsection*{CEI-ACAT Method}

Alternatively, we can combine the two statistics into a single
$p$-value measurement using the Aggregated Cauchy Association Test
(ACAT) \cite{LIU2019410}. ACAT merges multiple $p$-values by mapping
each one to a Cauchy distribution and then returning a single
$p$-value. We use this method to consolidate the two one-sided
$p$-values for $Z_{d_i}$ and $Z_{r_i}$ into a single score.  ACAT
remains valid even if dependency between the two indicators
exists. The ACAT method constructs the quantity

\begin{equation}
  T_i = \frac{1}{2}\tan\big[\pi\,(0.5 - p_{d_i})\big]
  + \frac{1}{2} \tan\big[\pi\,(0.5 - p_{r_i})\big]
\end{equation}
from the one-sided $p$-values $p_{d_i}=1-\Phi(Z_{d_i})$ and
$p_{r_i}=1-\Phi(Z_{r_i})$, respectively. The corresponding effective
$p$-value for clone $i$ is then calculated from
\begin{equation}
    p_i^{\text{ACAT}} = 0.5 - \frac{1}{\pi}\arctan(T_i).
\end{equation}
Again, BH-FDR control is utilized to filter out false
positive clones.


\section*{Results}
We now apply the methods described above to SARS-CoV-2 vaccination
data to identify vaccine-associated TCR sequences. The FDR will be
controlled by the BH procedure, and we set $\alpha=1\%-10\%$ as a
possible range for selection.

\subsection*{Comparison of positive-associated clones identified
  by different methods}

When applying a $10\%$ FDR to both the CEI-Joint and the CEI-ACAT
methods, the table shows strong inter-individual and inter-visit
heterogeneity in the number of positive clones. To provide a clear
overview of this variability, we list the number of
positive-associated clones in Table~\ref{tab:total}. The CEI-Joint
method typically identifies many more clones than the CEI-ACAT method
(for example, it reports thousands of clones for CLE0108 and CLE0117).
In contrast, CEI-ACAT tends to be more conservative, although it
occasionally calls more positive clones than CEI-Joint in specific
cases (such as for CLE0101 B–P1, CLE0100 B–P2, and CLE0136
B-P1). Additionally, the results patterns differ depending on the
visit pairs: some individuals show their strongest signals in P2
compared to P1 (indicating a later response), while others have the
largest signal in P2 compared to B (indicating an overall strong
post-vaccination expansion). Several participants, however, exhibited
very few positive clones across all comparisons (\textit{e.g.},
CLE0100 and CLE0136). Since subject CLE0099 received the chimpanzee
adenovirus-vectored vaccine, the set of activated TCR clones in this
individual consisted of clones responding not only to SARS-CoV-2
antigens but also to antigens derived from the adenoviral vector
itself.
  \begin{table*}[!t]
    \centering
    \renewcommand*{\arraystretch}{1.3}
    \begin{tabular}{c|c|c|c|cc|cc|cc}
    \hline
    \multirow{2}{*}{Individuals} & \multirow{2}{*}{$R_{\rm B}$} & \multirow{2}{*}{$R_{\rm P1}$} & \multirow{2}{*}{$R_{\rm P2}$} & \multicolumn{2}{c|}{B vs P1} & \multicolumn{2}{c|}{B vs P2} & \multicolumn{2}{c}{P1 vs P2} \\  \cline{5-10}
    \: & \: & \: & \: & \multicolumn{1}{c|}{Joint}  & \multicolumn{1}{c|}{ACAT} & \multicolumn{1}{c|}{Joint} & ACAT  & \multicolumn{1}{c|}{Joint} & ACAT  
    \\ \hline
    \multicolumn{1}{c|}{CLE0083} & {$184,988$} & \multicolumn{1}{c|}{$315,615$} & \multicolumn{1}{c|}{$322,569$}
    & \multicolumn{1}{c|}{$359$} & \multicolumn{1}{c|}{$5$} & \multicolumn{1}{c|}{$373$} & $3$  & \multicolumn{1}{c|}{$2$} & $2$  
    \\ \hline
    \multicolumn{1}{c|}{CLE0099} & {$223,828$} & \multicolumn{1}{c|}{$85,703$} & \multicolumn{1}{c|}{$231,395$}
    & \multicolumn{1}{c|}{$451$} & \multicolumn{1}{c|}{$216$} & \multicolumn{1}{c|}{$44$} & $27$  & \multicolumn{1}{c|}{$2584$} & $12$  
    \\ \hline
    \multicolumn{1}{c|}{CLE0100} & \multicolumn{1}{c|}{$245,481$} & \multicolumn{1}{c|}{$219,434$} & \multicolumn{1}{c|}{$267,635$}
    & \multicolumn{1}{c|}{$1$} & \multicolumn{1}{c|}{$1$}  & \multicolumn{1}{c|}{$4$} & \multicolumn{1}{c|}{$21$} & \multicolumn{1}{c|}{$17$} & $16$    
    \\ \hline
    \multicolumn{1}{c|}{CLE0101} & \multicolumn{1}{c|}{$252,373$} & \multicolumn{1}{c|}{$271,931$} & \multicolumn{1}{c|}{$177,847$}
    & \multicolumn{1}{c|}{$24$} & \multicolumn{1}{c|}{$89$} & \multicolumn{1}{c|}{$101$} & \multicolumn{1}{c|}{$47$} & \multicolumn{1}{c|}{$202$} & $33$ 
    \\ \hline
    \multicolumn{1}{c|}{CLE0108} & \multicolumn{1}{c|}{$136,808$} & \multicolumn{1}{c|}{$591,274$} & \multicolumn{1}{c|}{$208,447$}
    & \multicolumn{1}{c|}{$5127$} & \multicolumn{1}{c|}{$5$} & \multicolumn{1}{c|}{$46$} & \multicolumn{1}{c|}{$2$} & \multicolumn{1}{c|}{$2001$} & $60$ 
    \\ \hline
    \multicolumn{1}{c|}{CLE0117} & \multicolumn{1}{c|}{$133,666$} & \multicolumn{1}{c|}{$89,962$} & \multicolumn{1}{c|}{$369,441$}
    & \multicolumn{1}{c|}{$117$} & \multicolumn{1}{c|}{$5$}  & \multicolumn{1}{c|}{$4248$} & \multicolumn{1}{c|}{$3$} & \multicolumn{1}{c|}{$5119$} & $11$ 
    \\ \hline
    \multicolumn{1}{c|}{CLE0136} & \multicolumn{1}{c|}{$370,013$} & \multicolumn{1}{c|}{$386,065$} & \multicolumn{1}{c|}{$297,501$}
    & \multicolumn{1}{c|}{$2$} & \multicolumn{1}{c|}{$4$}  & \multicolumn{1}{c|}{$5$} & \multicolumn{1}{c|}{$6$} & \multicolumn{1}{c|}{$17$} & $7$ 
    \\ \hline
    \end{tabular}
    \caption{Table of the total number of positive-associated clones
      across multiple participants. $R_{\rm B}$, $R_{\rm P1}$, and
      $R_{\rm P2}$ are the number of clones (richness) detected for
      each individual in the sample B, P1, and P2, respectively. In
      the ${\rm B}$ versus ${\rm P1}$, ${\rm B}$ versus ${\rm P2}$,
      and ${\rm P1}$ versus ${\rm P2}$ columns, we list the number of
      SARS-CoV-2 clones identified through their sufficient increases
      in abundance as determined by our CEI-Joint and CEI-ACAT
      approaches. We set the FDR $\alpha=10\%$ for selection.}
    \label{tab:total}
\end{table*}

We compared the positively associated clones identified by our
CEI-Joint and CEI-ACAT methods, as illustrated in Fig. \ref{fig:venn2}
and in Figs. \ref{fig:1}-\ref{fig:6} in the Supplementary
Material. The positive clones selected at different times were
analyzed by comparing sample P1 with sample B, sample P2 with sample
B, and sample P2 with sample P1. Results are presented in
Fig.~\ref{fig:venn3} and Figs.~\ref{fig:7}-\ref{fig:12} in
the Supplementary Material. For better visualization, both figures
display selections made by the CEI-Joint method at a $10\%$ FDR and
selections made by the CEI-ACAT method at a $1\%$ FDR.
\begin{figure}[!h]
    \centering
    \includegraphics[width=4.9in]{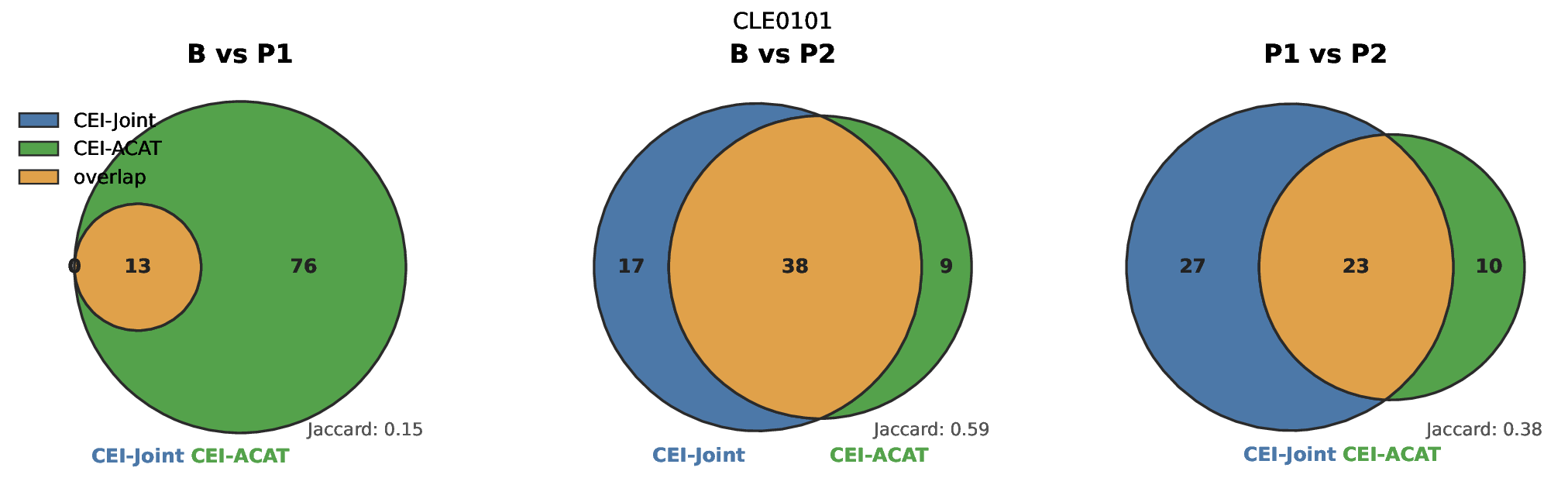}
     \caption{Comparisons between the positive sequences called by
       each method for individual CLE0101. The blue and green circles
       represent sets of clones identified by the CEI-Joint and
       CEI-ACAT methods, respectively. The size of each circle
       indicates the number of positive clones detected. The yellow
       portion illustrates the overlap of positive clones identified
       by both the CEI-Joint and CEI-ACAT methods. Left: Clones
       identified through comparison of sample P1 to sample B. Middle:
       Clones identified through comparison of sample P2 to sample
       B. Right: Clones identified through comparison of sample P2 to
       sample P1.}
    \label{fig:venn2}
\end{figure}
\begin{figure}[!h]
    \centering
    \includegraphics[width=3.2in]{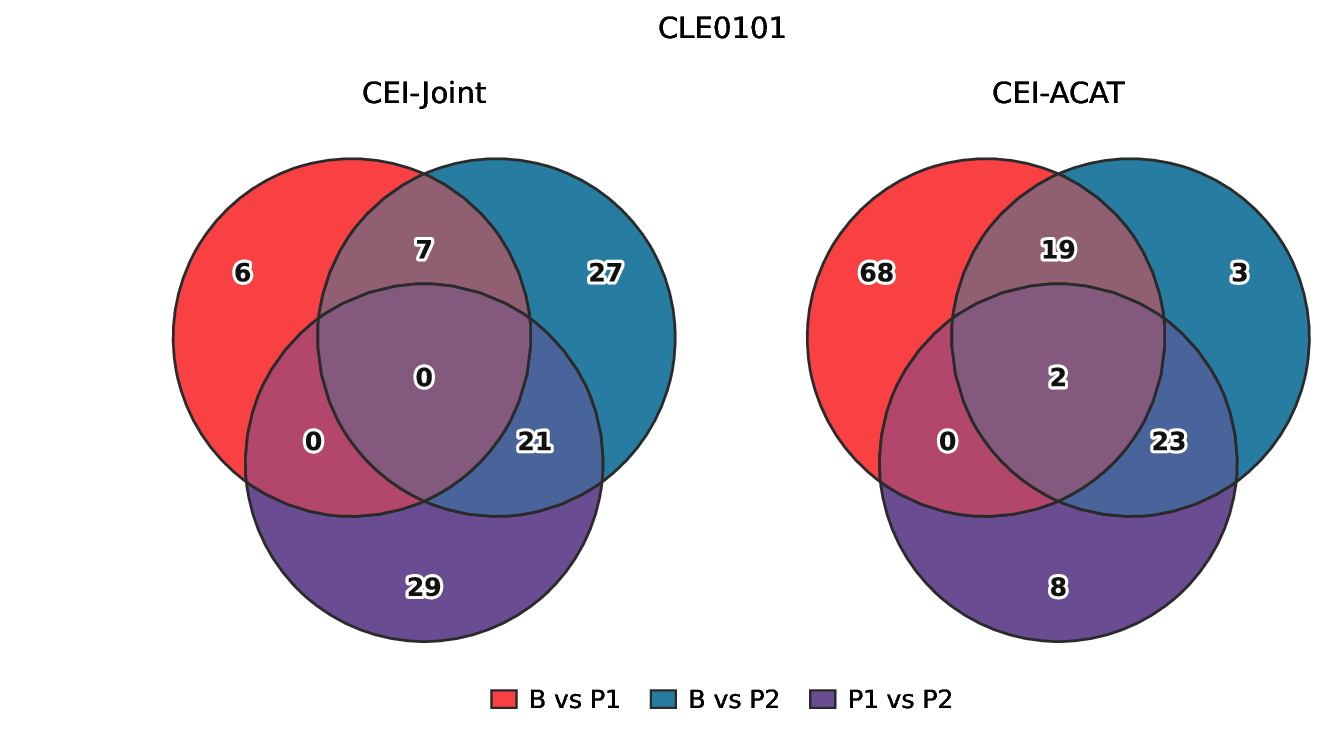}
     \caption{Comparisons between the positive sequences sampled at
       different times for subject CLE0101. The subplots, left and
       right, represent comparisons using the CEI-Joint and CEI-ACAT
       methods, respectively. Red, blue, and purple circles correspond
       to positive clones identified by comparing sample P1 with B, P2
       with B, and P2 with P1. The size of each circle reflects the
       number of positive clones identified. Left: Clones identified
       via the CEI-Joint method. Right: Clones identified by the
       CEI-ACAT method.}
    \label{fig:venn3}
\end{figure}

\subsubsection*{Comparison with standard heuristic selection criteria (Adap) and edgeR}

Our TCR identification methods were further evaluated by comparing
their predictions with those from methods previously developed to
identify differentially abundant clones \cite{dewitt,rojas2023} in
different contexts.

The standard heuristic selection method (Adap) previously used by
Adaptive Biotechnologies designates a clone as activated if its
sampled baseline and post-vaccination counts exhibit at least a 2-fold
increase \textit{and} that the abundances pass a two-sided binomial
test with a $p$-value below $0.01$. In this analysis, the two-sided
binomial test p-values are adjusted using the BH procedure to
effectively control the FDR. Furthermore, to ensure the reliability of
the detection, only clones with a minimum count abundance of five in
the sample are considered. edgeR, which tests negative binomial or
generalized linear models, was also applied to our single sample (no
replicates) data.



\begin{figure*}[!h]
    \begin{center}
        \includegraphics[width=5.4in]{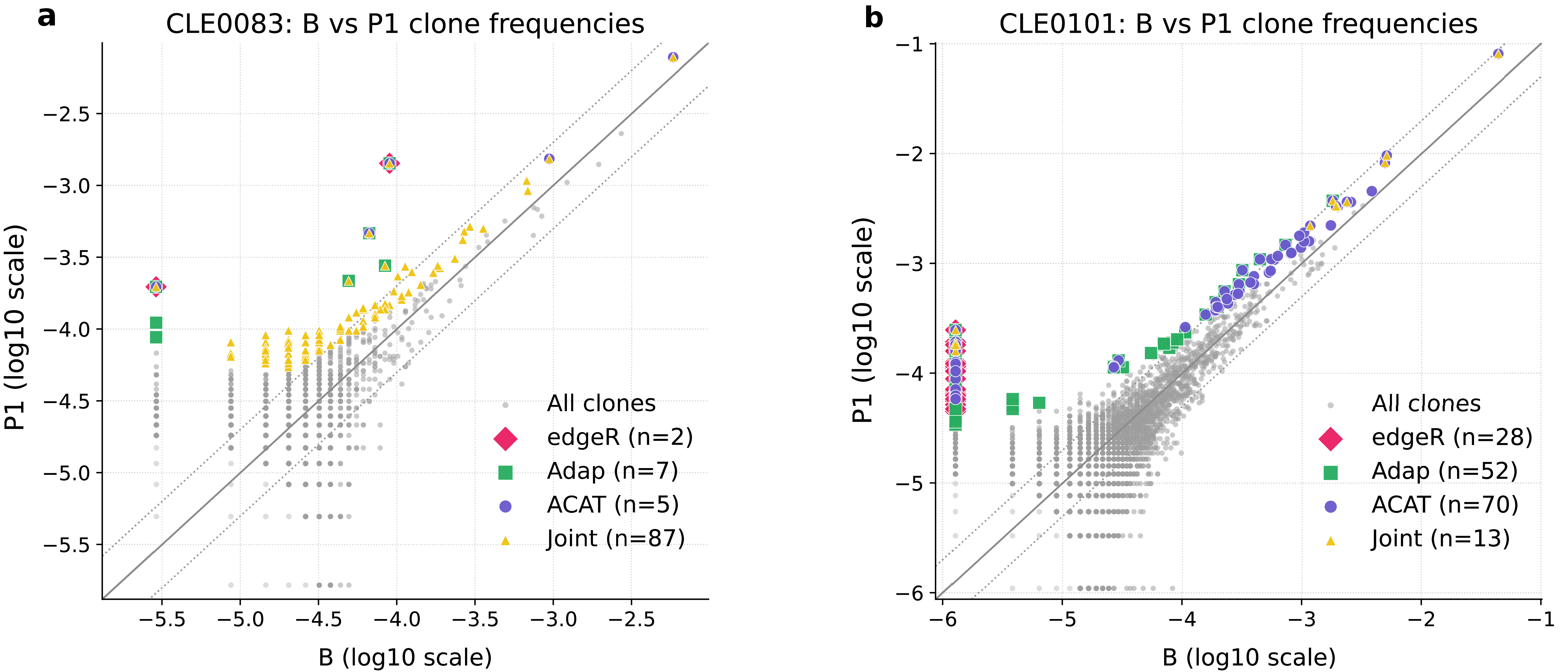}
    \end{center}
    \caption{Abundances of individual TCR clones before vaccination
      (B) are plotted along the x-axis, while their abundances after
      vaccination (P1) are plotted along the y-axis. (a) and (b)
      correspond to subjects CLE0083 and CLE0101, respectively. Clones
      that are statistically significantly expanded (blue) are
      identified by a corrected binomial model \cite{stats} and
      additionally, a threshold of at least a 2-fold increase or
      decrease from B to P1.  Each dot corresponds to a unique TCR
      clone: grey circles represent robust clones with an abundance
      greater than $5$ templates, while clones with fewer than $5$
      templates are shown in white to indicate they fall below the
      threshold for reliable detection. Purple squares represent the
      expanded clones called by our ACAT method using a FDR at $1\%$,
      while green circles denote the expanded clones identified by the
      standard heuristic approach. Similarly, clones that have been
      detected by edgeR are indicated by the red triangles. These
      results illustrate the overlap between the clones identified by
      the ACAT method, the clones detected using the standard
      heuristic criteria, and those called by edgeR for differential
      abundance analysis.}
    \label{compare}
\end{figure*}

Figure~\ref{compare} compares the clones identified using two
heuristic criteria (Adap and edgeR) with those identified using our
CEI-Joint and CEI-ACAT methods for the subjects CLE0083 and
CLE0101. All methods were controlled at a $1\%$ FDR. The panels
highlight significant subject-specific differences and distinct
profiles for each method.

For CLE0083 (a), our CEI-Joint test identifies a considerable number
of expansions ($n=87$), far exceeding the counts for CEI-ACAT ($n=5$),
Adap ($n=7$), and edgeR ($n=2$). Some of the CEI-Joint calls lie just
above the B=P1 diagonal, indicating a broad but modest shift following
vaccination. In contrast, for CLE0101 (b), the rankings
shift. CEI-ACAT ($n=70$) and Adap ($n=52$) identify the largest sets
along the rising trend, while edgeR falls in between with $n=28$. The
CEI-Joint method is the most stringent, identifying only $n=13$
clones.

Across both subjects, selected clones predominantly sit above the
diagonal (indicating expansion), but the methods emphasize different
regions of the plot. The standard heuristic method focuses solely on
larger log-fold changes, while edgeR tends to favor high log-fold
change signals with low clone abundance at baseline. Our CEI-ACAT and
CEI-Joint methods strike a balance between absolute differences and
log-fold changes. This results in many calls when there is a diffuse,
concordant shift (as seen in CLE0083), but fewer calls when changes
cluster along a tight trajectory (as seen in CLE0101).

Our collection of statistical techniques provides a more systematic
framework for fine-tuning FDR, which allows us to control the
stringency in identifying activated clones. These flexible tools can
be customized to detect population expansions in different biological
contexts, ensuring that the identification process is both rigorous
and adaptable to varying experimental conditions.

\section*{Discussion and Conclusions}

Our study builds on the efforts to map TCRs to SARS-CoV-2 antigens by
offering a high-resolution, longitudinal view of both CD4$^{+}$ and
CD8$^{+}$ T cell responses at three critical timepoints:
pre-vaccination, post-dose 1, and post-dose 2. This design allows for
precise tracking of TCR clonal expansion and repertoire reshaping in
response to antigenic stimulation. By focusing on COVID-naive
individuals, this study avoids confounding from prior infection and
provides a clean baseline for understanding vaccine-induced T cell
immunity—an essential component of long-term protection and immune
memory.

To determine whether changes in TCR abundances are deemed
vaccination-sensitive (\textit{e.g.}, whether associated T cells
proliferate following vaccination) requires us to carefully define
statistical thresholds. We presented several measures to detect clones
whose abundances increased significantly after vaccination. Our
framework quantifies each clone with two complementary statistics—the
rescaled abundance change of a clone $d_{i}$ and its log-fold change
in abundance $r_{i}$.

\begin{figure}[htp!]
    \centering
    \includegraphics[width=0.98\linewidth]{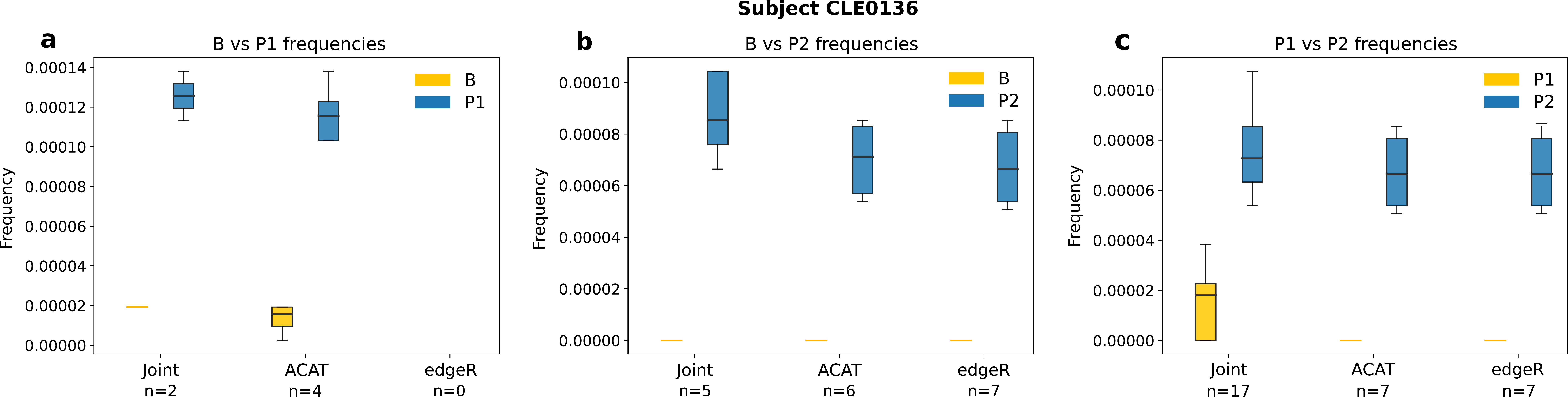}
    \caption{For subject CLE0136, the boxplots compare clone
      frequencies between time points for the clones called by each
      method ($n$ listed below panels): (a) B vs P1, (b) B vs P2, and
      (c) P1 vs P2. For each contrast, boxplots display the
      frequencies of the identified clones by comparing two samples
      using three different methods: CEI-Joint, CEI-ACAT, and edgeR
      (with the number of calls, $n$, under each method). The
      CEI-Joint method yields significantly higher frequencies at both
      baseline and post-vaccination samples. Note that edgeR calls
      no positive sequences any significant expansion in the B vs P1
      comparison, even when we relaxed the FDR.}
    \label{fig:edgeR}
\end{figure}

Human samples make collection of replicates challenging and make it
difficult to filter out experimental noise during the selection
process and sequencing procedure. To validate our methods, we compared
our selections with the simpler rule ``$r_i \ge 2$'' (plus a minimum
of five template counts), NoisET and edgeR. NoisET and edgeR are
well-known packages designed for clonal expansion detection. NoisET
explicitly learns an experimental noise model and then applies a
Dirichlet-Multinomial Bayesian test for expansion under
stimulation. However, our dataset lacks replicates, which prevents us
from using NoisET to learn about the noise and detect expanded
clones. On the other hand, edgeR uses empirical Bayes moderation
techniques for scenarios with very few or no replicates. However, it
is too conservative and highly sensitive to the abundance of clones in
the baseline sample, such that it can only detect expanded clones if
their abundance in the baseline sample is nearly zero. If the size of
these clones is even slightly larger in the baseline sample, edgeR may
fail to identify the expansion. We provide a case example for
reference in Fig.~\ref{fig:edgeR}.

\begin{figure}[htp!]
    \centering
    \includegraphics[width=0.97\linewidth]{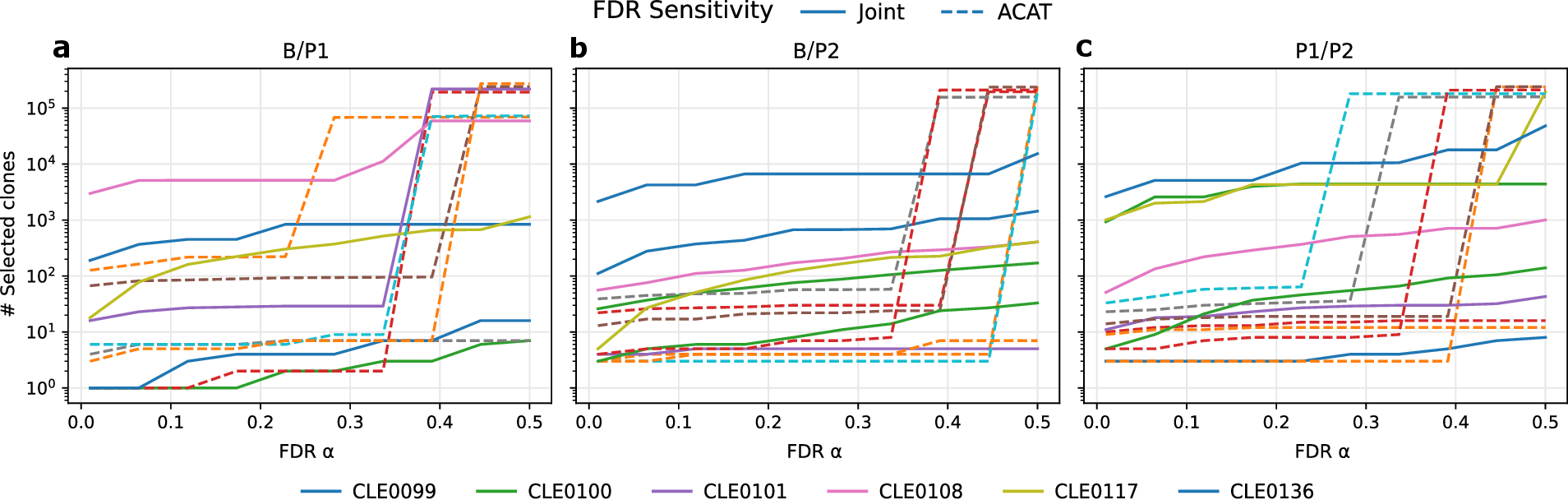}
    \caption{Sensitivity of the number of positive clones selected
      with respect to the FDR threshold $\alpha$, where $\alpha \in
      (0, 0.5]$. The dashed lines correspond to the CEI-ACAT method,
    while the solid lines correspond to the CEI-Joint method. (a)
    Sensitivity analysis conducted by comparing sample P1 with B. (b)
    Sensitivity analysis conducted by comparing sample P2 with B. (c)
    sensitivity analysis conducted by comparing sample P2 with P1.}
    \label{fig:sensitivity}
\end{figure}

We performed sensitivity analysis of the number of positive clones
selected by scanning the FDR $\alpha$ from $0.001$ to $0.5$. Since
most clones are of small abundance with relatively small changes, the
number of detected clones increases sharply once $\alpha$ exceeds a
certain level, shown in the Fig.~\ref{fig:sensitivity}.


Our general methods can be applied to any multispecies system subject
to an external perturbation that yields before and after snapshots of
abundances. Using the same metrics and thresholds also enable
direct comparisons across different induction protocols and
experimental contexts, allowing one to quantify overlaps in results
from different experimental/clinical methods. For example, using
changes in the \textit{in vivo} T-cell abundance as an indicator of
vaccination response led to significantly fewer identified TCRs than
using cellular immune assays using antigen libraries \citep{MIRA}. The
substantial difference between clones that were measured to have
expanded \textit{in vivo} and those identified through antigen
recognition in cell culture suggests a more complex pathway from
antigen recognition to T cell amplification and carries important
clinical implications. Besides overcounting low-affinity TCRs,
discrepancies between the \textit{in vivo} and \textit{ex vivo}
methods may arise from other causes such as finite sampling from
individuals, different timescales between \textit{ex vivo} assays and
\textit{in vivo} clone abundance dynamics, and dynamical fluctuations
of interacting T-cell populations in an individual. How these effects
influence changes in clonal populations subjected to different
stimulation protocols will be quantitatively compared in future
investigations that use the methods described here.

\vspace{5mm}
\section*{Declaration of Interest Statement}

HP, BB, SAB are employees of Adaptive Biotechnologies and hold stock
or stock options in the company. YP, TC, and OY do not have any
commercial or financial relationships with Adaptive Biotechnologies.

\section*{CRediT author contribution statement}

YP: Writing – review \& editing, Writing – original draft,
Investigation, Conceptualization, Data curation, Methodology,
Visualization, Formal Analysis; CH: Writing – review \& editing,
Investigation, Data curation, Resources; BB, HP, and SB: Data
curation, Visualization, Resources, Formal Analysis, Review \&
Editing; TC: Writing – Original Draft, Writing – Review \& Editing,
Visualization, Supervision, Methodology, Conceptualization; OY:
Writing – Review \& Editing, Conceptualization, Methodology,
Supervision.


\section*{Acknowledgements}
The authors are grateful for support from the UCLA-CDU CFAR grant
no. AI152501.  Additional funding was provided by private
philanthropic donors (including William Moses, Mari Edelman, Beth
Friedman, Dana and Matt Walden, Kathleen Poncher, Scott Z. Burns,
Gwyneth Paltrow and Brad Falchuk), with additional infrastructure
support from the UCLA-Charles Drew University Center for AIDS Research
(NIH grant AI152501), James B. Pendleton Trust, and McCarthy
Foundation. Support for TCR immunosequencing was provided by Adaptive Biotechnologies, Seattle, WA, USA.


\section*{Supplementary materials}
Additional comparisons of identified clones using different methods,
across all other subjects are provided in the Supplementary materials
available in the online version.


\section*{Data availability}
Post-processed immunosequencing data (as described here) are freely
available at the ImmuneACCESS database,
\url{https://clients.adaptivebiotech.com/immuneaccess}, and will also
be provided upon request.

\section*{Code availability}

The software package is freely available under the MIT license from the GitHub website, \url{https://github.com/ybpan16/CEI}.



\newpage

\bibliographystyle{elsarticle-num}
\bibliography{ref}

@article{rojas2023,
  title={{Personalized RNA neoantigen vaccines stimulate T cells in
                  pancreatic cancer}},
  author={Rojas, Luis A and Sethna, Zachary and Soares, Kevin C and
                  Olcese, Cristina and Pang, Nan and Patterson, Erin
                  and Lihm, Jayon and Ceglia, Nicholas and Guasp,
                  Pablo and Chu, Alexander and others},
  journal={Nature},
  volume={618},
  number={7963},
  pages={144--150},
  year={2023},
  publisher={Nature Publishing Group UK London}
}

@article{dewitt,
author={W. S. DeWitt and R. O. Emerson and P. Lindau and others},
title={Dynamics of the cytotoxic {T} cell response to a model of acute
                  viral infection},
journal={Journal of Virology},
year={2015},
volume={89},
pages={4517--4526}}

@article{stats,
author={Y. Benjamini and Y. Gavrilov},
title={A simple forward selection procedure based on false discovery
                  rate control},
journal={Annals of Applied Statistics},
volume={3},
number={1},
pages={179--198},
year={2009}}

@article{MIRA,
  title={{Multiplex Identification of Antigen-Specific T Cell
                  Receptors Using a Combination of Immune Assays and
                  Immune Receptor Sequencing}},
  author={Mark Klinger and Francois Pepin and Jen Wilkins and Thomas
                  Asbury and Tobias Wittkop and Jianbiao Zheng and
                  Martin Moorhead and Malek Faham},
  journal={PLoS ONE},
  volume={10},
  number={10},
  pages={e0141561},
  year={2015}}

@article{Baden2021mrna,
    author = {Baden, Lindsey R. and El Sahly, Hana M. and Essink,
                  Brandon and Kotloff, Karen and Frey, Sharon and
                  Novak, Rick and Diemert, David and Spector, Stephen
                  A. and Rouphael, Nadine and Creech, C. Buddy and
                  McGettigan, John and Khetan, Shishir and Segall,
                  Nathan and Solis, Joel and Brosz, Adam and Fierro,
                  Carlos and Schwartz, Howard and Neuzil, Kathleen and
                  Corey, Lawrence and Gilbert, Peter and Janes, Holly
                  and Follmann, Dean and Marovich, Mary and Mascola,
                  John and Polakowski, Laura and Ledgerwood, Julie and
                  Graham, Barney S. and Bennett, Hamilton and Pajon,
                  Rolando and Knightly, Conor and Leav, Brett and
                  Deng, Weiping and Zhou, Honghong and Han, Shu and
                  Ivarsson, Melanie and Miller, Jacqueline and Zaks,
                  Tal},
    title = {Efficacy and Safety of the m{RNA}-1273 {SARS}-{C}o{V}-2 Vaccine},
    journal = {New England Journal of Medicine},
    volume = {384},
    number = {5},
    pages = {403-416},
    year = {2021},
    doi = {10.1056/NEJMoa2035389},
    note ={PMID: 33378609},
    URL = { https://doi.org/10.1056/NEJMoa2035389},
    eprint = {https://doi.org/10.1056/NEJMoa2035389}
}

@article{garcia2021multiple,
    title={{Multiple SARS-CoV-2 variants escape neutralization by
                  vaccine-induced humoral immunity}},
    author={Garcia-Beltran, Wilfredo F. and Lam, Evan C. and Denis,
                  Kerri St. and Nitido, Adam D. and Garcia, Zeidy H. and
                  Hauser, Blake M. and Feldman, Jared and Pavlovic,
                  Maia N. and Gregory, David J. and Poznansky, Mark C.
                  and others},
    journal={Cell},
    volume={184},
    number={9},
    pages={2372--2383},
    year={2021},
    publisher={Elsevier}
}

@article{goel2021mrna,
  title={{mRNA vaccines induce durable immune memory to SARS-CoV-2 and
                  variants of concern}},
  author={Goel, Rishi R. and Painter, Mark M. and Apostolidis,
                  Sokratis A. and Mathew, Divij and Meng, Wenzhao and
                  Rosenfeld, Aaron M. and Lundgreen, Kendall A. and
                  Reynaldi, Arnold and Khoury, David S. and Pattekar,
                  Ajinkya and others},
  journal={Science},
  volume={374},
  number={6572},
  pages={abm0829},
  year={2021},
  publisher={American Association for the Advancement of Science}
}

@article{ibarrondo2021primary,
  title={{Primary, recall, and decay kinetics of SARS-CoV-2 vaccine
                  antibody responses}},
  author={Ibarrondo, F. Javier and Hofmann, Christian and Fulcher,
                  Jennifer A. and Goodman-Meza, David and Mu, William
                  and Hausner, Mary Ann and Ali, Ayub and Balamurugan,
                  Arumugam and Taus, Ellie and Elliott, Julie and
                  others},
  journal={ACS Nano},
  volume={15},
  number={7},
  pages={11180--11191},
  year={2021},
  publisher={ACS Publications}
}

@article{link2023estimation,
  title={{Estimation of COVID-19 mRNA vaccine effectiveness and
                  COVID-19 illness and severity by vaccination status
                  during Omicron BA. 4 and BA. 5 sublineage periods}},
  author={Link-Gelles, Ruth and Levy, Matthew E. and Natarajan, Karthik
                  and Reese, Sarah E. and Naleway, Allison L. and
                  Grannis, Shaun J. and Klein, Nicola P and DeSilva,
                  Malini B. and Ong, Toan C. and Gaglani, Manjusha and
                  others},
  journal={JAMA network open},
  volume={6},
  number={3},
  pages={e232598--e232598},
  year={2023},
  publisher={American Medical Association}
}

@article{polack2020safety,
  title={{Safety and efficacy of the BNT162b2 mRNA Covid-19 vaccine}},
  author={Polack, Fernando P. and Thomas, Stephen J. and Kitchin,
                  Nicholas and Absalon, Judith and Gurtman, Alejandra
                  and Lockhart, Stephen and Perez, John L. and
                  P{\'e}rez Marc, Gonzalo and Moreira, Edson D. and
                  Zerbini, Cristiano and others},
  journal={New England Journal of Medicine},
  volume={383},
  number={27},
  pages={2603--2615},
  year={2020},
  publisher={Mass Medical Soc}
}

@article{zhang2022humoral,
  title={Humoral and cellular immune memory to four {COVID-19} vaccines},
  author={Zhang, Zeli and Mateus, Jose and Coelho, Camila H. and Dan,
                  Jennifer M. and Moderbacher, Carolyn Rydyznski and
                  G{\'a}lvez, Rosa Isela and Cortes, Fernanda H. and
                  Grifoni, Alba and Tarke, Alison and Chang, James and
                  others},
  journal={Cell},
  volume={185},
  number={14},
  pages={2434--2451},
  year={2022},
  publisher={Elsevier}
}

@article{augusto2023common,
  title={A common allele of {HLA} is associated with asymptomatic
                  {SARS-CoV-2} infection},
  author={Augusto, Danillo G. and Murdolo, Lawton D. and
                  Chatzileontiadou, Demetra S. .M and Sabatino Jr., Joseph
                  J. and Yusufali, Tasneem and Peyser, Noah D. and
                  Butcher, Xochitl and Kizer, Kerry and Guthrie,
                  Karoline and Murray, Victoria W. and others},
  journal={Nature},
  volume={620},
  number={7972},
  pages={128--136},
  year={2023},
  publisher={Nature Publishing Group UK London}
}

@article{buckley2022hla,
  title={{HLA-dependent variation in SARS-CoV-2 CD8+ T cell
                  cross-reactivity with human coronaviruses}},
  author={Buckley, Paul R. and Lee, Chloe H. and Pereira Pinho, Mariana
                  and Ottakandathil Babu, Rosana and Woo, Jeongmin and
                  Antanaviciute, Agne and Simmons, Alison and Ogg,
                  Graham and Koohy, Hashem},
  journal={Immunology},
  volume={166},
  number={1},
  pages={78--103},
  year={2022},
  publisher={Wiley Online Library}
}

@article{sette2023t,
  title={{T cell responses to SARS-CoV-2}},
  author={Sette, Alessandro and Sidney, John and Crotty, Shane},
  journal={Annual Review of Immunology},
  volume={41},
  pages={343--373},
  year={2023},
  publisher={Annual Reviews}
}

@article{taus2023persistent,
  title={{Persistent memory despite rapid contraction of circulating T
                  cell responses to SARS-CoV-2 mRNA vaccination}},
  author={Taus, Ellie and Hofmann, Christian and Ibarrondo, F. Javier
                  and Fulcher, Jennifer A. and Kitchen, Scott G. and
                  Tobin, Nicole H. and Yang, Otto O.},
  journal={Frontiers in Immunology},
  volume={14},
  pages={1100594},
  year={2023},
  publisher={Frontiers}
}

@article{taus2022dominant,
  title={{Dominant CD8+ T cell nucleocapsid targeting in SARS-CoV-2
                  infection and broad spike targeting from
                  vaccination}},
  author={Taus, Ellie and Hofmann, Christian and Ibarrondo, Francisco
                  Javier and Hausner, Mary Ann and Fulcher, Jennifer A.
                  and Krogstad, Paul and Ferbas, Kathie G and Tobin,
                  Nicole H. and Rimoin, Anne W. and Aldrovandi, Grace M.
                  and others},
  journal={Frontiers in Immunology},
  volume={13},
  pages={835830},
  year={2022},
  publisher={Frontiers Media SA}
}

@book{murphy2016janeway,
  title={Janeway's Immunobiology},
  author={Murphy, Kenneth and Weaver, Casey},
  year={2016},
  publisher={Garland Science}
}

@article{nikolich2004many,
  title={The many important facets of {T}-cell repertoire diversity},
  author={Nikolich-{\v{Z}}ugich, Janko and Slifka, Mark K. and
                  Messaoudi, Ilhem},
  journal={Nature Reviews Immunology},
  volume={4},
  number={2},
  pages={123--132},
  year={2004},
  publisher={Nature Publishing Group UK London}
}

@article{sewell2012must,
  title={Why must {T} cells be cross-reactive?},
  author={Sewell, Andrew K.},
  journal={Nature Reviews Immunology},
  volume={12},
  number={9},
  pages={669--677},
  year={2012},
  publisher={Nature Publishing Group UK London}
}

@article{arstila1999direct,
  title={A direct estimate of the human $\alpha$$\beta$ {T} cell
                  receptor diversity},
  author={Arstila, T. Petteri and Casrouge, Armanda and Baron,
                  V{\'e}ronique and Even, Jos and Kanellopoulos, Jean
                  and Kourilsky, Philippe},
  journal={Science},
  volume={286},
  number={5441},
  pages={958--961},
  year={1999},
  publisher={American Association for the Advancement of Science}
}

@article{lythe2016many,
  title={How many {TCR} clonotypes does a body maintain?},
  author={Lythe, Grant and Callard, Robin E. and Hoare, Rollo L. and
                  Molina-Par{\'\i}s, Carmen},
  journal={Journal of Theoretical Biology},
  volume={389},
  pages={214--224},
  year={2016},
  publisher={Elsevier}
}

@article{mora2019many,
  title={How many different clonotypes do immune repertoires contain?},
  author={Mora, Thierry and Walczak, Aleksandra M.},
  journal={Current Opinion in Systems Biology},
  volume={18},
  pages={104--110},
  year={2019},
  publisher={Elsevier}
}

@article{qi2014diversity,
  title={Diversity and clonal selection in the human {T}-cell
                  repertoire},
  author={Qi, Qian and Liu, Yi and Cheng, Yong and Glanville, Jacob
                  and Zhang, David and Lee, Ji-Yeun and Olshen,
                  Richard A. and Weyand, Cornelia M. and Boyd, Scott D.
                  and Goronzy, J{\"o}rg J.},
  journal={Proceedings of the National Academy of Sciences},
  volume={111},
  number={36},
  pages={13139--13144},
  year={2014},
  publisher={National Acad Sciences}
}

@article{warren2011exhaustive,
  title={Exhaustive {T}-cell repertoire sequencing of human peripheral
                  blood samples reveals signatures of antigen
                  selection and a directly measured repertoire size of
                  at least 1 million clonotypes},
  author={Warren, Ren{\'e} L. and Freeman, J. Douglas and Zeng, Thomas
                  and Choe, Gina and Munro, Sarah and Moore, Richard
                  and Webb, John R. and Holt, Robert A.},
  journal={Genome Research},
  volume={21},
  number={5},
  pages={790--797},
  year={2011},
  publisher={Cold Spring Harbor Lab}
}

@article{jenkins2009composition,
  title={On the composition of the preimmune repertoire of {T} cells
                  specific for peptide--major histocompatibility
                  complex ligands},
  author={Jenkins, Marc K. and Chu, H. Hamlet and McLachlan, James
                  B. and Moon, James J.},
  journal={Annual Review of Immunology},
  volume={28},
  pages={275--294},
  year={2009},
  publisher={Annual Reviews}
}

@article{aoki2024cd8+,
  title={{CD8+ T cell memory induced by successive SARS-CoV-2 mRNA
                  vaccinations is characterized by shifts in clonal
                  dominance}},
  author={Aoki, Hiroyasu and Kitabatake, Masahiro and Abe, Haruka and
                  Xu, Peng and Tsunoda, Mikiya and Shichino, Shigeyuki
                  and Hara, Atsushi and Ouji-Sageshima, Noriko and
                  Motozono, Chihiro and Ito, Toshihiro and others},
  journal={Cell Reports},
  year={2024},
  publisher={Elsevier}
}

@article{mallajosyula2021cd8+,
  title={{CD8+ T cells specific for conserved coronavirus epitopes
                  correlate with milder disease in patients with
                  COVID-19}},
  author={Mallajosyula, Vamsee and Ganjavi, Conner and Chakraborty,
                  Saborni and McSween, Alana M. and Pavlovitch-Bedzyk,
                  Ana Jimena and Wilhelmy, Julie and Nau, Allison and
                  Manohar, Monali and Nadeau, Kari C. and Davis, Mark
                  M.},
  journal={Science Immunology},
  volume={6},
  number={61},
  pages={eabg5669},
  year={2021},
  publisher={American Association for the Advancement of Science}
}

@article{swadling2022pre,
  title={{Pre-existing polymerase-specific T cells expand in abortive
                  seronegative SARS-CoV-2}},
  author={Swadling, Leo and Diniz, Mariana O. and Schmidt, Nathalie M.
                  and Amin, Oliver E. and Chandran, Aneesh and Shaw,
                  Emily and Pade, Corinna and Gibbons, Joseph M. and Le
                  Bert, Nina and Tan, Anthony T. and others},
  journal={Nature},
  volume={601},
  number={7891},
  pages={110--117},
  year={2022},
  publisher={Nature Publishing Group UK London}
}

@article{ibarrondo2020rapid,
  title={{Rapid decay of anti--SARS-CoV-2 antibodies in persons with
                  mild COVID-19}},
  author={Ibarrondo, F. Javier and Fulcher, Jennifer A. and
                  Goodman-Meza, David and Elliott, Julie and Hofmann,
                  Christian and Hausner, Mary A. and Ferbas, Kathie G.
                  and Tobin, Nicole H. and Aldrovandi, Grace M. and
                  Yang, Otto O.},
  journal={New England Journal of Medicine},
  volume={383},
  number={11},
  pages={1085--1087},
  year={2020},
  publisher={Mass Medical Soc}
}

@article{rodda2021functional,
  title={{Functional SARS-CoV-2-specific immune memory persists after
                  mild COVID-19}},
  author={Rodda, Lauren B. and Netland, Jason and Shehata, Laila and
                  Pruner, Kurt B. and Morawski, Peter A. and Thouvenel,
                  Christopher D. and Takehara, Kennidy K. and
                  Eggenberger, Julie and Hemann, Emily A. and Waterman,
                  Hayley R. and others},
  journal={Cell},
  volume={184},
  number={1},
  pages={169--183},
  year={2021},
  publisher={Elsevier}
}

@article{wheatley2020evolution,
  title={Evolution of immunity to {SARS-CoV-2} in mild-moderate {COVID}-19},
  author={Wheatley, Adam K. and Juno, Jennifer A. and Wang, Jing J. and
                  Selva, Kevin J. and Reynaldi, Arnold and Tan,
                  Hyon-Xhi and Shi Lee, Wen and Wragg, Kathleen M. and
                  Kelly, Hannah G. and Esterbauer, Robyn and others},
  journal={Nature Communications},
  volume={12},
  pages={116},
  year={2021}
}

@article{jokinen2023tcrconv,
  title={TCRconv: predicting recognition between {T} cell receptors and
                  epitopes using contextualized motifs},
  author={Jokinen, Emmi and Dumitrescu, Alexandru and Huuhtanen, Jani
                  and Gligorijevi{\'c}, Vladimir and Mustjoki, Satu
                  and Bonneau, Richard and Heinonen, Markus and
                  L{\"a}hdesm{\"a}ki, Harri},
  journal={Bioinformatics},
  volume={39},
  number={1},
  pages={btac788},
  year={2023},
  publisher={Oxford University Press}
}

@article{ferretti2020unbiased,
  title={{Unbiased screens show CD8+ T cells of COVID-19 patients
                  recognize shared epitopes in SARS-CoV-2 that largely
                  reside outside the spike protein}},
  author={Ferretti, Andrew P. and Kula, Tomasz and Wang, Yifan and
                  Nguyen, Dalena M. V. and Weinheimer, Adam and Dunlap,
                  Garrett S. and Xu, Qikai and Nabilsi, Nancy and
                  Perullo, Candace R. and Cristofaro, Alexander W. and
                  others},
  journal={Immunity},
  volume={53},
  number={5},
  pages={1095--1107},
  year={2020},
  publisher={Elsevier}
}

@article{grifoni2021sars,
  title={{SARS-CoV-2 human T cell epitopes: Adaptive immune response
                  against COVID-19}},
  author={Grifoni, Alba and Sidney, John and Vita, Randi and Peters,
                  Bjoern and Crotty, Shane and Weiskopf, Daniela and
                  Sette, Alessandro},
  journal={Cell Host \& Microbe},
  volume={29},
  number={7},
  pages={1076--1092},
  year={2021},
  publisher={Elsevier}
}

@article{grifoni2020targets,
  title={{Targets of T cell responses to SARS-CoV-2 coronavirus in
                  humans with COVID-19 disease and unexposed
                  individuals}},
  author={Grifoni, Alba and Weiskopf, Daniela and Ramirez, Sydney I.
                  and Mateus, Jose and Dan, Jennifer M. and
                  Moderbacher, Carolyn Rydyznski and Rawlings, Stephen
                  A. and Sutherland, Aaron and Premkumar, Lakshmanane
                  and Jadi, Ramesh S. and others},
  journal={Cell},
  volume={181},
  number={7},
  pages={1489--1501},
  year={2020},
  publisher={Elsevier}
}

@article{Kuchenbecker2015IMSEQ,
    author = {Kuchenbecker, Leon and Nienen, Mikalai and Hecht, Jochen
                  and Neumann, Avidan U. and Babel, Nina and Reinert,
                  Knut and Robinson, Peter N.}, 
    title = {{IMSEQ—a fast and error aware approach to immunogenetic
                  sequence analysis}},
    volume = {31}, 
    number = {18}, 
    pages = {2963-2971}, 
    year = {2015}, 
    doi = {10.1093/bioinformatics/btv309}, 
    URL = {http://bioinformatics.oxfordjournals.org/content/31/18/2963.abstract}, 
    eprint = {http://bioinformatics.oxfordjournals.org/content/31/18/2963.full.pdf+html}, 
    journal = {Bioinformatics} 
}

@article{KL1951div,
    author = {S. Kullback and R. A. Leibler},
    title = {{On Information and Sufficiency}},
    volume = {22},
    journal = {The Annals of Mathematical Statistics},
    number = {1},
    publisher = {Institute of Mathematical Statistics},
    pages = {79 -- 86},
    year = {1951},
    doi = {10.1214/aoms/1177729694},
    URL = {https://doi.org/10.1214/aoms/1177729694}
}

@article{Kantorovich1960mathematical,
    title={Mathematical methods of organizing and planning production},
    author={Kantorovich, Leonid V.},
    journal={Management Science},
    volume={6},
    number={4},
    pages={366--422},
    year={1960},
    publisher={INFORMS}
}

@article{Vaserstein1969markov,
  title={Markov processes over denumerable products of spaces,
                  describing large systems of automata},
  author={Vaserstein, Leonid Nisonovich},
  journal={Problemy Peredachi Informatsii},
  volume={5},
  number={3},
  pages={64--72},
  year={1969},
  publisher={Russian Academy of Sciences, Branch of Informatics,
                  Computer Equipment and~…}
}

@article{Singhal2001modern,
  title={Modern information retrieval: A brief overview},
  author={Singhal, Amit and others},
  journal={IEEE Data Eng. Bull.},
  volume={24},
  number={4},
  pages={35--43},
  year={2001}
}

@article{Bhattacharyya1946measure,
  title={On a measure of divergence between two multinomial
                  populations},
  author={Bhattacharyya, Anil},
  journal={Sankhy{\=a}: The Indian Journal of Statistics},
  pages={401--406},
  year={1946},
  publisher={JSTOR}
}

@article{koraichi2022noiset,
  title={{NoisET: noise learning and expansion detection of T-cell receptors}},
  author={Koraichi, Meriem Bensouda and Touzel, Maximilian Puelma and
                  Mazzolini, Andrea and Mora, Thierry and Walczak,
                  Aleksandra M.},
  journal={The Journal of Physical Chemistry A},
  volume={126},
  number={40},
  pages={7407--7414},
  year={2022},
  publisher={ACS Publications}
}

@article{chen2018differential,
  title={Differential methylation analysis of reduced representation
                  bisulfite sequencing experiments using edge{R}},
  author={Chen, Yunshun and Pal, Bhupinder and Visvader, Jane E. and
                  Smyth, Gordon K.},
  journal={F1000Research},
  volume={6},
  pages={2055},
  year={2018}
}

@article{LIU2019410,
    title = {{ACAT}: A Fast and Powerful p-Value Combination Method for Rare-Variant Analysis in Sequencing Studies},
    journal = {The American Journal of Human Genetics},
    volume = {104},
    number = {3},
    pages = {410-421},
    year = {2019},
    issn = {0002-9297},
    doi = {https://doi.org/10.1016/j.ajhg.2019.01.002},
    url = {https://www.sciencedirect.com/science/article/pii/S0002929719300023},
    author = {Yaowu Liu and Sixing Chen and Zilin Li and Alanna C. Morrison and Eric Boerwinkle and Xihong Lin}
}



\newpage
\setcounter{figure}{0}
    \renewcommand{\thefigure}{S\arabic{figure}}
\begin{center}
  \Large{\textbf{Supplementary Material}}
\end{center}





\subsection*{Variance for $\log_2 p$}

Let $s \sim \text{Poisson}(\lambda)$ where $\lambda = Sp$ and the real proportion $f = \frac{s}{S}$. Then, we have
\begin{equation}
    {\mathbb{E}}\hat{p}  = \lambda \quad\text{and}\quad {\rm Var}(\hat{p} ) = \lambda
\end{equation}
Given the function $g(p) = \log_2 p$, we apply the Taylor expansion to find
\begin{equation}
    g(f) = g(p) + g'(p)(f -p) + O\left((f-p)^2\right)
\end{equation}
Hence,
\begin{equation}
    {\rm Var}(g(f)) \approx {\rm Var}\left(g(p) + g'(p)(f -p)\right) = \left(g'(p)\right)^2 {\rm Var}(f) = \frac{1}{(\ln 2)^2 p^2} \cdot \frac{pS}{S^2} = \frac{1}{\ln 2} \cdot \frac{1}{p S}
\end{equation}

\subsection*{Comparisons between positive sequences called by each method for different individuals}

Comparisons between the positive sequences called by each method for
each subject. The blue and green circles represent sets of clones
identified by the CEI-Joint and CEI-ACAT methods, respectively. The
size of each circle indicates the number of positive clones
detected. The yellow portion illustrates the overlap of positive
clones identified by both CEI-Joint and CEI-ACAT methods. Left: Clones
identified through comparison of sample P1 with sample B. Middle:
Clones identified through comparison of sample P2 with sample
B. Right: Clones identified through comparison of sample P2 with
sample P1.
\begin{figure}[htbp]
    \begin{center}
        \includegraphics[width=5.7in]{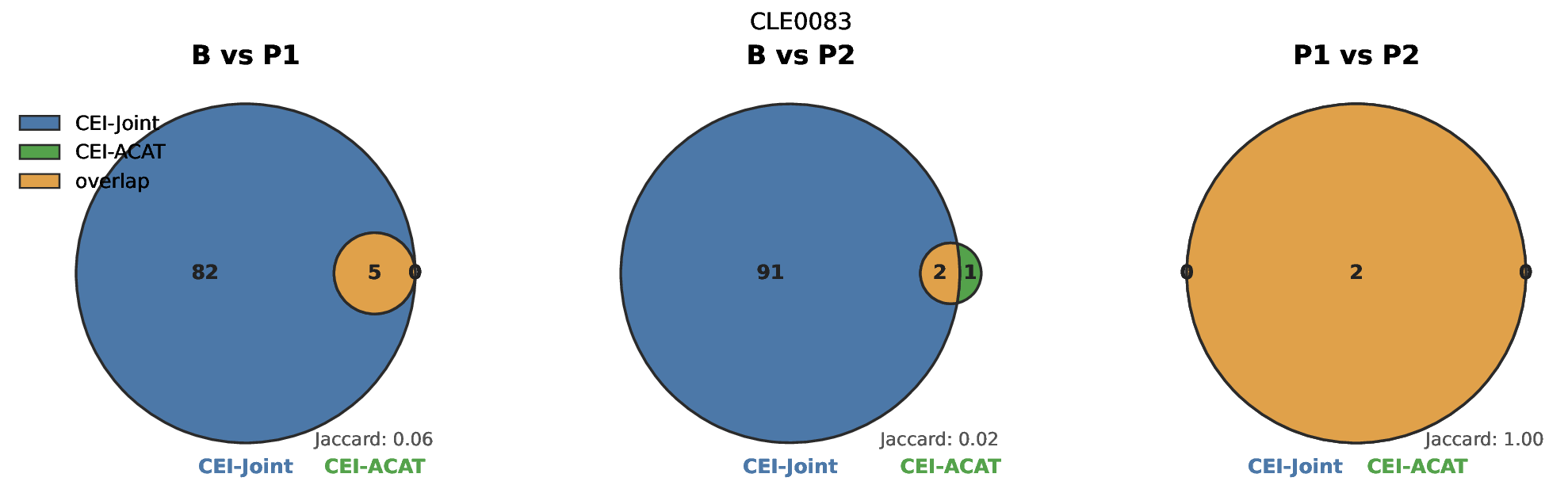}
    \end{center}
    \caption{Comparisons between the positive sequences called by each
      method for individual CLE0083.}
    \label{fig:1}
\end{figure}
\begin{figure}[htbp]
    \begin{center}
        \includegraphics[width=5.7in]{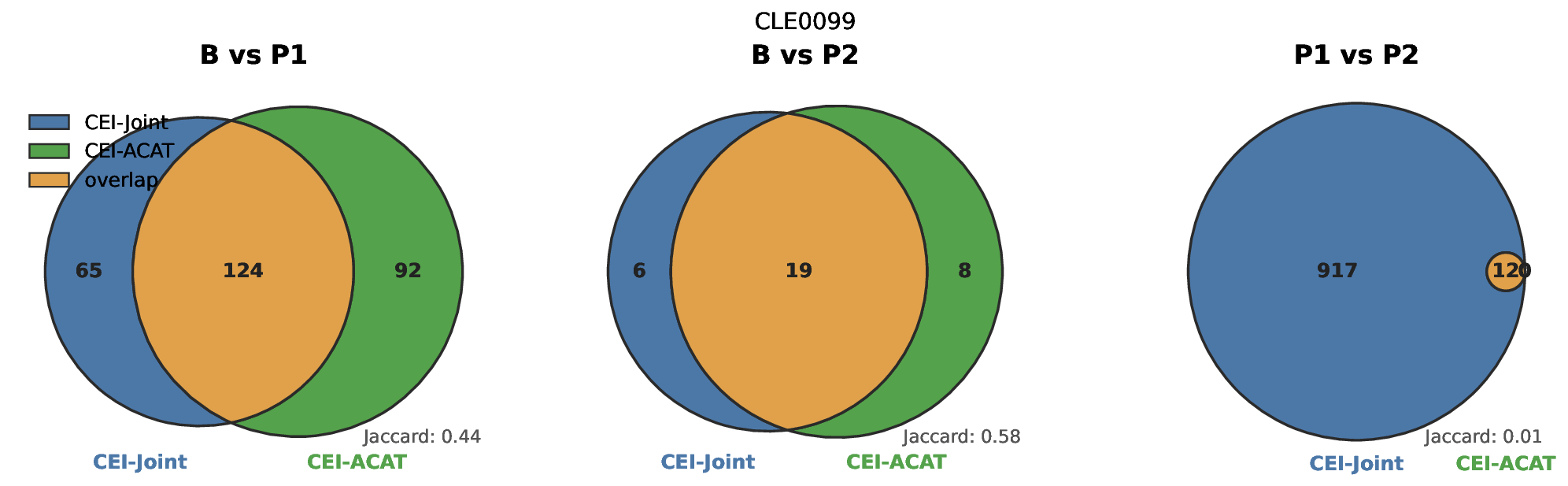}
    \end{center}
    \caption{Comparisons between the positive sequences called by each
      method for individual CLE0099.}
    \label{fig:2}
\end{figure}
\begin{figure}[htbp]
    \begin{center}
        \includegraphics[width=5.7in]{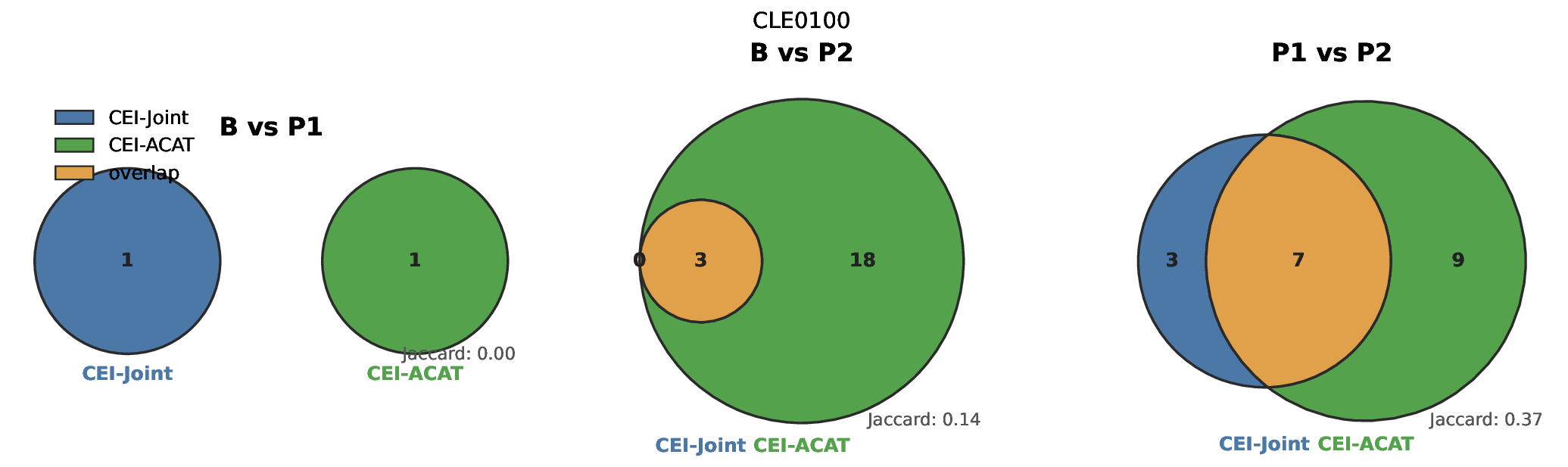}
    \end{center}
    \caption{Comparisons between the positive sequences called by each
      method for individual CLE0100.}
    \label{fig:3}
\end{figure}
\begin{figure}[htbp]
    \begin{center}
        \includegraphics[width=5.7in]{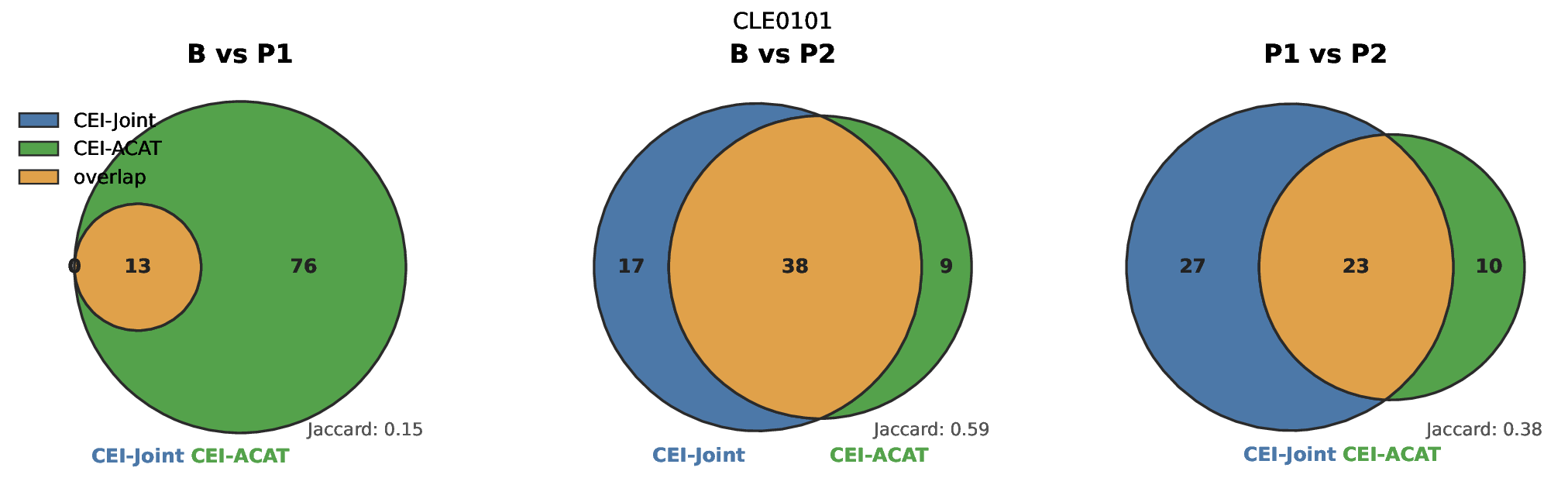}
    \end{center}
    \caption{Comparisons between the positive sequences called by each
      method for individual CLE0101.}
    \label{fig:4}
\end{figure}
\begin{figure}[htbp]
    \begin{center}
        \includegraphics[width=5.7in]{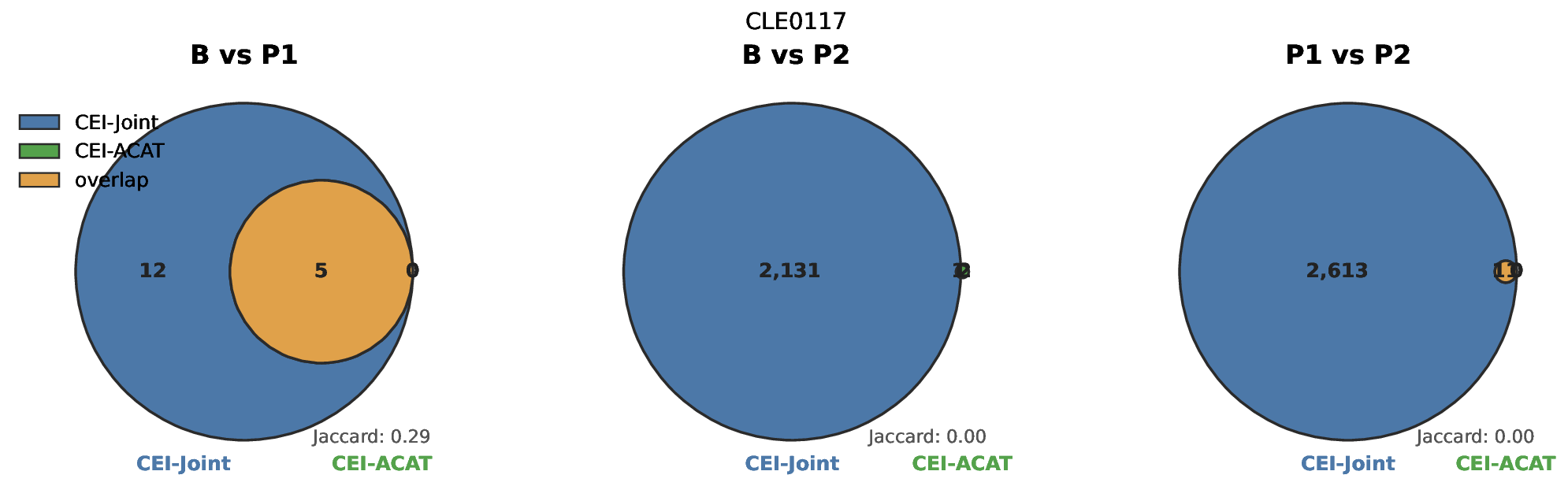}
    \end{center}
    \caption{Comparisons between the positive sequences called by each
      method for individual CLE0117.}
    \label{fig:5}
\end{figure}
\begin{figure}[htbp]
    \begin{center}
        \includegraphics[width=5.7in]{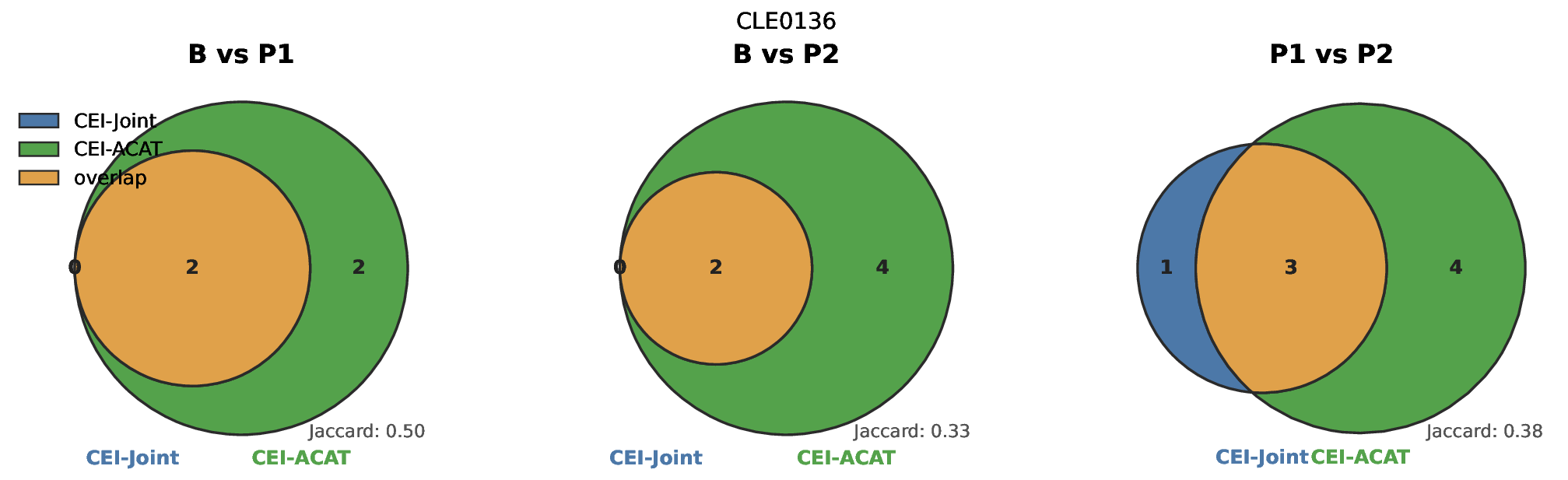}
    \end{center}
    \caption{Comparisons between the positive sequences called by each
      method for individual CLE0136.}
    \label{fig:6}
\end{figure}

\newpage

\subsection*{Comparisons between positive sequences sampled at different times for different individuals}

Comparisons between the positive sequences sampled at different times
for different individuals. The subplots, left and right, represent
comparisons using the CEI-Joint and CEI-ACAT methods,
respectively. Red, blue, and purple circles correspond to positive
clones identified by comparing sample P1 with B, P2 with B, and P2
with P1. The size of each circle reflects the number of positive
clones identified. Left: Clones identified via the CEI-Joint
method. Right: Clones identified by the CEI-ACAT method.
\begin{figure}[htbp]
    \begin{center}
        \includegraphics[width=5.7in]{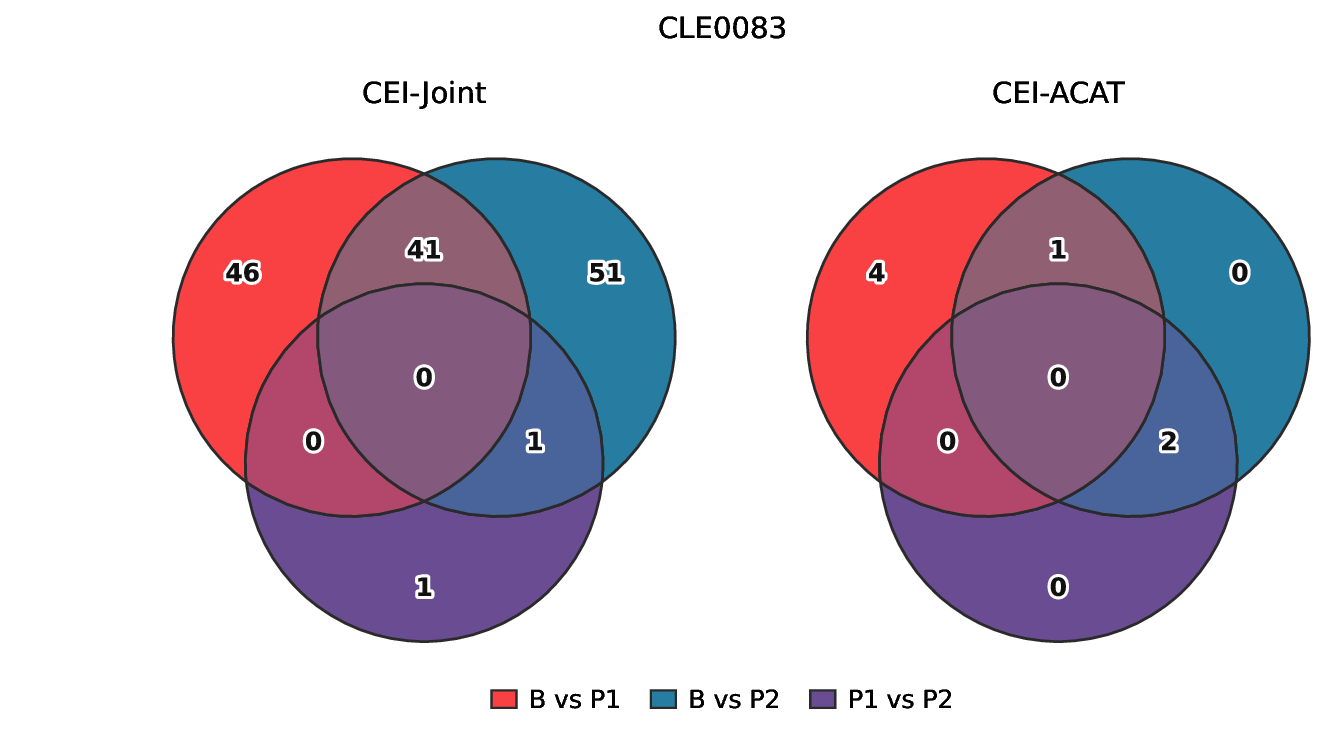}
    \end{center}
    \vspace{-2mm}
    \caption{Comparisons between the positive sequences sampled at different times for individual CLE0083}
    \label{fig:7}
\end{figure}
\begin{figure}[htbp]
    \begin{center}
        \includegraphics[width=5.7in]{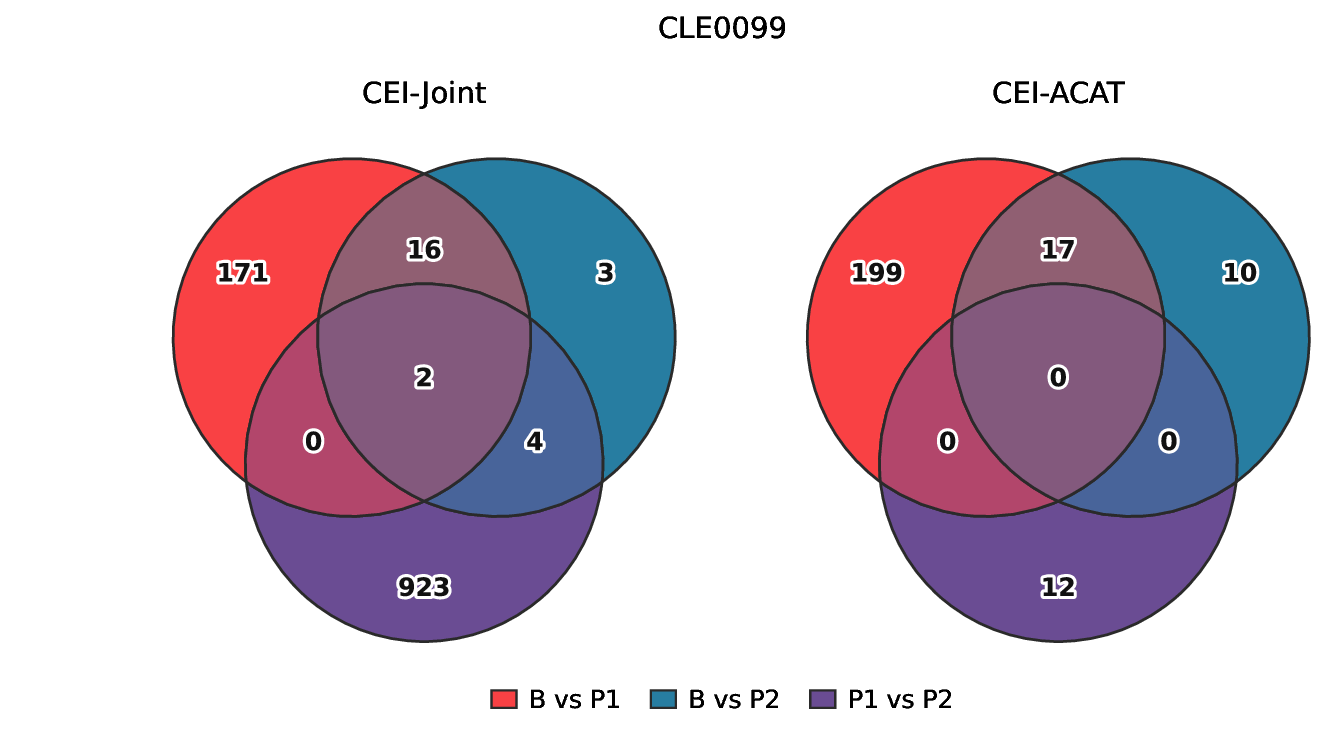}
    \end{center}
    \vspace{-2mm}
    \caption{Comparisons between the positive sequences sampled at different times for individual CLE0099}
    \label{fig:8}
\end{figure}
\begin{figure}[htbp]
    \begin{center}
        \includegraphics[width=5.7in]{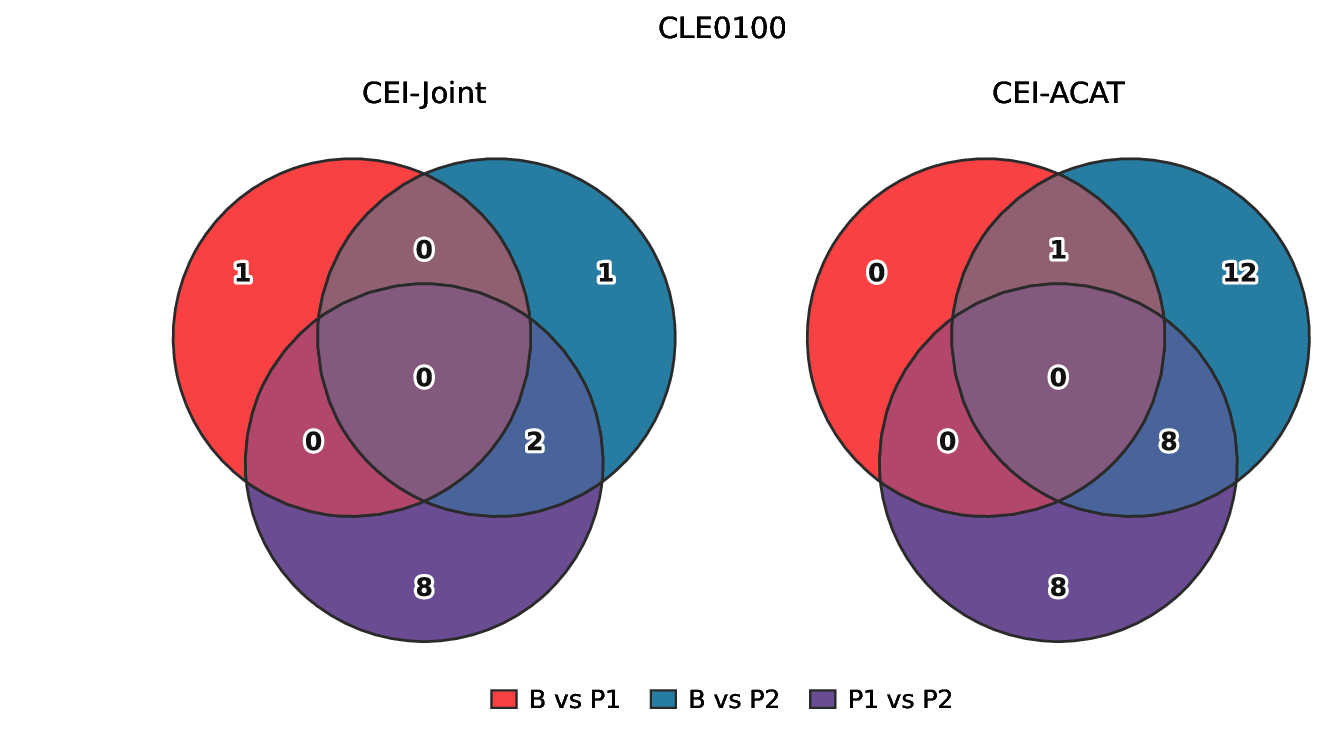}
    \end{center}
    \vspace{-2mm}
    \caption{Comparisons between the positive sequences sampled at different times for individual CLE0100}
    \label{fig:9}
\end{figure}
\begin{figure}[htbp]
    \begin{center}
        \includegraphics[width=5.7in]{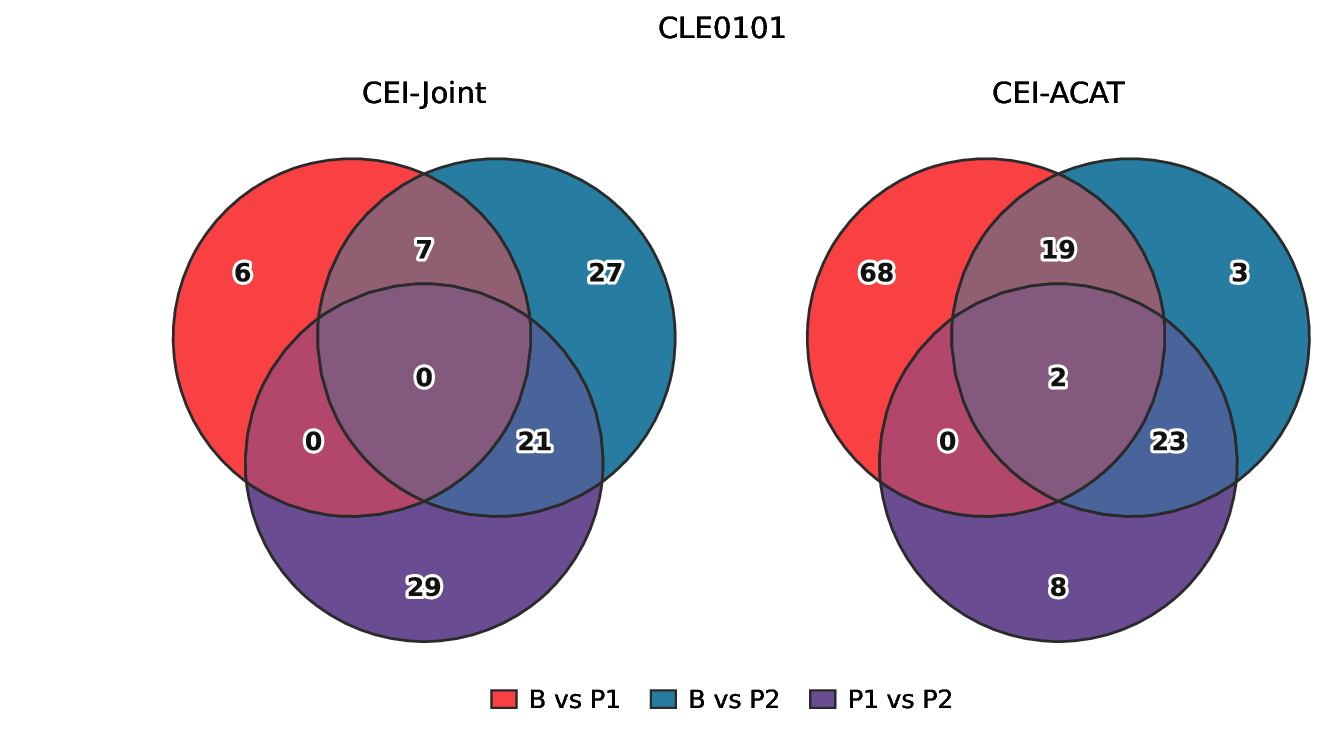}
    \end{center}
    \vspace{-2mm}
    \caption{Comparisons between the positive sequences sampled at different times for individual CLE0101}
    \label{fig:10}
\end{figure}
\begin{figure}[htbp]
    \begin{center}
        \includegraphics[width=5.7in]{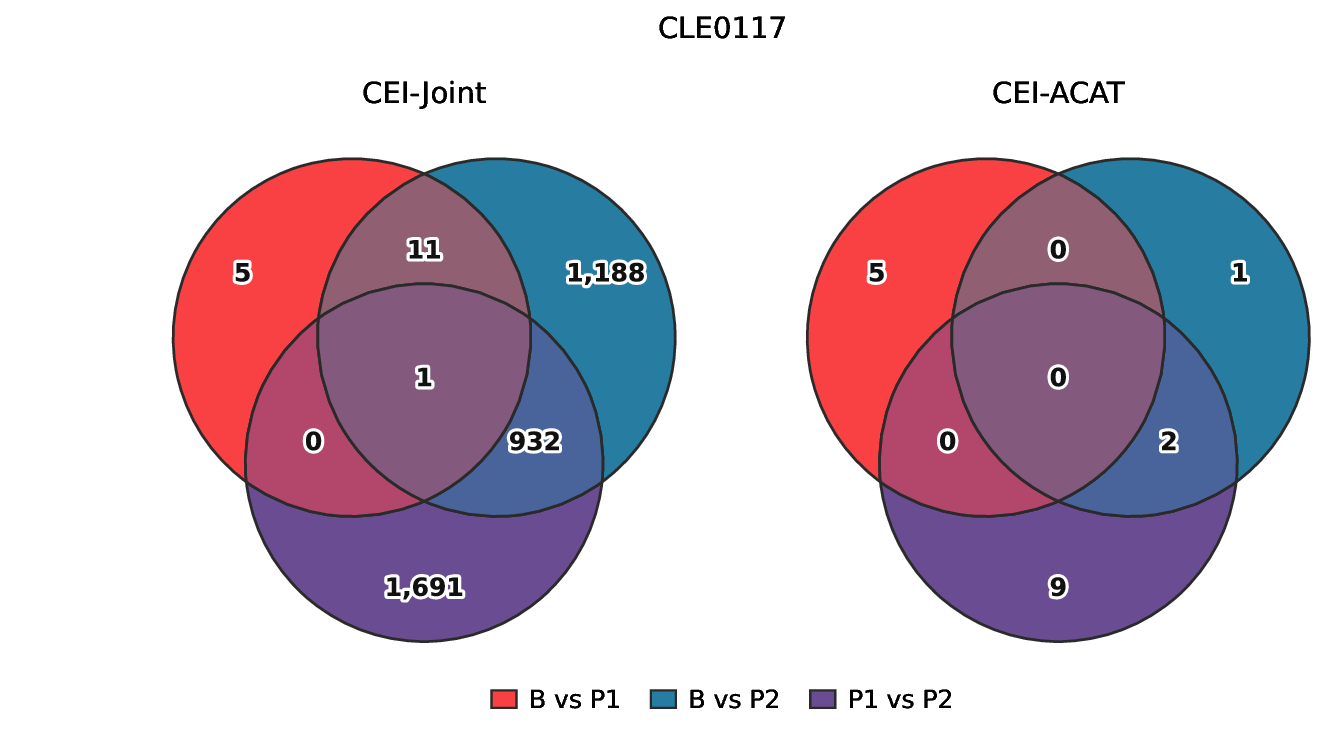}
    \end{center}
    \vspace{-2mm}
    \caption{Comparisons between the positive sequences sampled at different times for individual CLE0117}
    \label{fig:11}
\end{figure}
\begin{figure}[htbp]
    \begin{center}
        \includegraphics[width=5.7in]{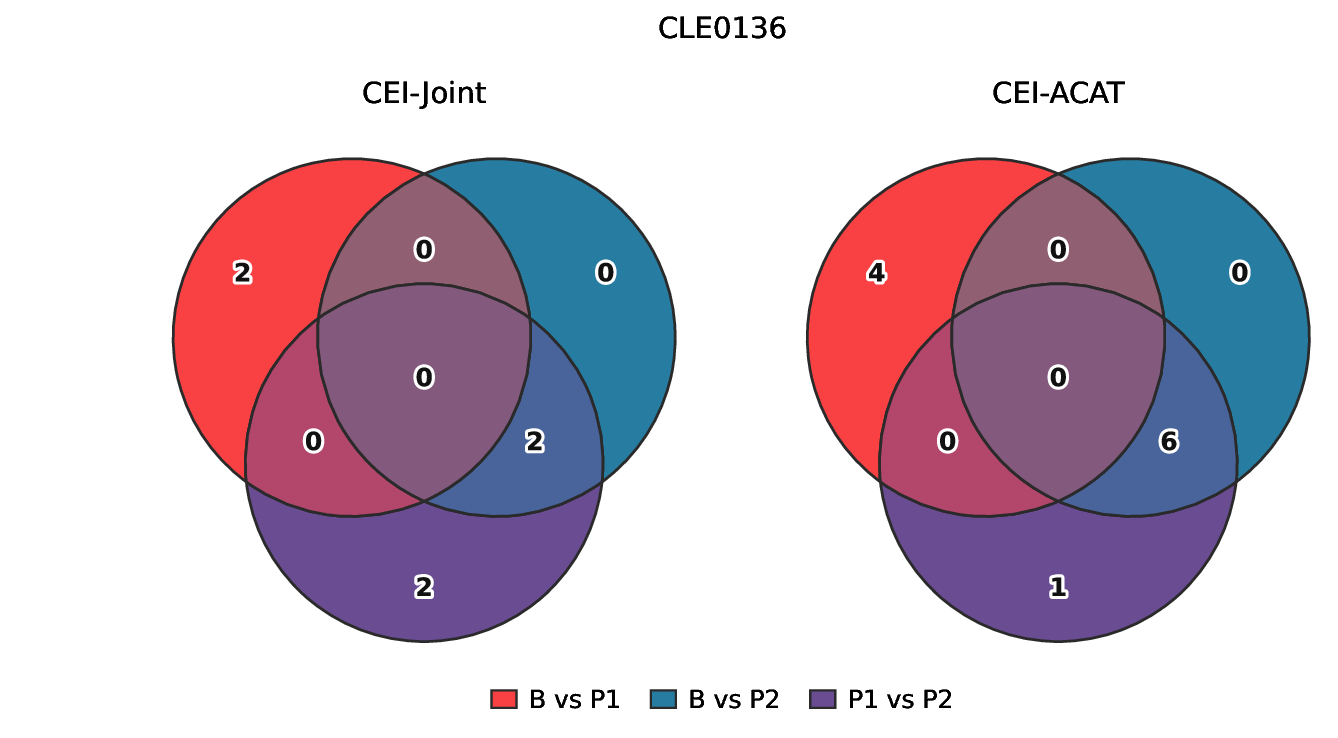}
    \end{center}
    \vspace{-2mm}
    \caption{Comparisons between the positive sequences sampled at different times for individual CLE0136}
    \label{fig:12}
\end{figure}

\end{document}